\documentclass[prx,aps,epsf,twocolumn,showpacs,superscriptaddress,longbibliography]{revtex4-1}
\usepackage[pdftex]{graphicx}
\usepackage{dcolumn}
\usepackage{bm}
\usepackage{epsfig}
\usepackage{latexsym} 
\usepackage{amsmath}
\usepackage{amssymb}
\usepackage{color}
\usepackage{array}
\usepackage{bbm}
\usepackage{hyperref}
\usepackage{color}
\usepackage{array}
\usepackage{cancel}
\usepackage{ulem}

\usepackage{booktabs}
\newcolumntype{C}[1]{>{\centering\arraybackslash}m{#1}}
\AtBeginDocument{
\heavyrulewidth=.08em
\lightrulewidth=.05em
\cmidrulewidth=.03em
\belowrulesep=.65ex
\belowbottomsep=0pt
\aboverulesep=.4ex
\abovetopsep=0pt
\cmidrulesep=\doublerulesep
\cmidrulekern=.5em
\defaultaddspace=.5em
}
\usepackage{booktabs}
\newcolumntype{R}[1]{>{\raggedleft\arraybackslash}p{#1}}
\AtBeginDocument{
\heavyrulewidth=.08em
\lightrulewidth=.05em
\cmidrulewidth=.03em
\belowrulesep=.65ex
\belowbottomsep=0pt
\aboverulesep=.4ex
\abovetopsep=0pt
\cmidrulesep=\doublerulesep
\cmidrulekern=.5em
\defaultaddspace=.5em
}

\newcommand{\<}{\langle}
\newcommand{\e}{\varepsilon}

\renewcommand{\>}{\rangle}
\renewcommand{\(}{\left(}
\renewcommand{\)}{\right)}
\renewcommand{\[}{\left[}
\renewcommand{\]}{\right]}
\renewcommand{\v}[1]{\mathbf{#1}} 

\renewcommand{\d}{\partial}

\newcommand{\eps}{\epsilon}

\newcommand{\Z}{\mathbb{Z}}
\newcommand{\T}{\mathcal{T}}

\begin{document}
\title{Classification of interacting topological Floquet phases in one dimension}
\author{Andrew C Potter}
\author{Takahiro Morimoto}
\author{Ashvin Vishwanath}
\affiliation{Department of Physics, University of California, Berkeley, CA 94720, USA}
\begin{abstract}
Periodic driving of a quantum system can enable new topological phases with no analog in static systems. In this paper we systematically classify one-dimensional topological and symmetry-protected topological (SPT) phases in interacting fermionic and bosonic quantum systems subject to periodic driving, which we dub Floquet SPTs (FSPTs). For physical realizations of interacting FSPTs, many-body localization by disorder is a crucial ingredient, required to obtain a stable phase that does not catastrophically heat to infinite temperature. We demonstrate that bosonic and fermionic FSPTs phases are classified by the same criteria as equilibrium phases, but with an enlarged symmetry group $\tilde G$, that now includes discrete time translation symmetry associated with the Floquet evolution. In particular, 1D bosonic FSPTs are classified by  projective representations of the enlarged symmetry group $H^2({\tilde G},U(1))$. We construct explicit lattice models for a variety of systems, and  then formalize the classification to demonstrate the completeness of this construction.  We also derive general constraints on localization and symmetry based on the representation theory of the symmetry group, and show that symmetry-preserving localized phases are possible only for Abelian symmetry groups. In particular, this rules out the possibility of many-body localized SPTs with continuous spin symmetry.
\end{abstract}
\maketitle

\section{Introduction}
Periodic driving of a quantum system enables one to tailor new interactions and achieve interesting quantum phases of matter. Such Floquet engineering has lead to various applications in quantum optical contexts, such as the engineering of artificial gauge fields\cite{Dalibard11}, as well as in solid-state contexts, e.g. to produce new Floquet-Bloch band structures\cite{Gu11,Wang13}, or understand non-linear optical phenomena\cite{Morimoto15}. In addition to providing new tools to engineer phases that could arise as ground-states of a different static Hamiltonian, periodic driving also opens up the possibility of engineering entirely new phases with no equilibrium analog.\cite{Kitagawa10,Jiang11,Rudner13,Asboth14,Gannot15,Thakurathi13,Thakurathi14,Iadecola15,Carpentier15} In the context of non-interacting particles, various examples of new topological phases that arise from driving are known, including dynamical Floquet analogs of Majorana fermions in 1D\cite{Jiang11}, and phases with chiral edge modes but vanishing Chern number in 2D\cite{Kitagawa10,Rudner13}. 

Heretofore, such investigations were largely restricted to non-interacting systems, as persistent driving of a generic interacting many-body system typically leads to catastrophic runaway heating towards a featureless infinite temperature steady state, for which there are no sharp notions of distinct phases. While very rapid driving, with frequency much larger than the natural interaction scales of the Hamiltonian can postpone this runaway heating for exponentially long times\cite{Abanin15}, new topological phases that occur exclusively in driven systems can be realized only in moderate frequency regimes where clean systems would be susceptible to heating issues.\cite{Kitagawa10,Jiang11,Rudner13,Asboth14,Gannot15} 

However, many-body localized (MBL) systems\cite{MBLReview} retain sharp spectral lines for local operators\cite{Nandkishore14}, and can therefore avoid energy absorption from off-resonant driving by a local Hamiltonian.\cite{Abanin14,Ponte15,Lazarides2015} Interestingly, despite being strongly localized, MBL systems can still exhibit non-trivial topological and SPT order.\cite{Huse13,Bauer13,Bahri15,Chandran14} This raises the general conceptual question of: which zero-temperature quantum phases can occur in the highly excited states of MBL systems? for which there is a growing systematic understanding.\cite{MBLSPT} The stability of MBL to Floquet driving enables sharp distinctions between dynamical phases of periodically driven matter\cite{Khemani15}, and extends this line of inquiry, and raises the prospect of realizing, not only familiar ground-state orders, but also fundamentally new interacting dynamical topological phases arising from driving. In this paper, we develop a systematic understanding of the structure of topological and symmetry protected topological (SPT) phases of periodically driven Floquet systems in one spatial dimension. 

Following Refs~\cite{Vosk13,Serbyn13,Bauer13,MBLSPT}, we begin by formulating a sharp criterion for many-body localizability in terms of the existence of an appropriate set of quasilocal conservation laws. We then study of interacting Floquet topological phases of fermions in 1D. After reviewing some ideas and explicit models for non-interacting Fermion Floquet SPTs\cite{Jiang11,Asboth14,Gannot15,Khemani15}, we then address the modification of the fermionic Floquet SPT classification due to interactions in all of the non-trivial classes of the 10-fold way\cite{Ryu10,Kitaev09}.  In the absence of interactions, periodic driving raises the new possibility of obtaining topologically protected edge modes with quasi-energy $\pi$,\cite{Kitagawa10,Jiang11} in addition to those with zero quasi-energy that are familiar from non-driven equilibrium systems.\cite{Kitaev01} As for the equilibrium SPTs, we find that interactions generally tend to reduce the set of nontrivial phases when the non-interacting classification contains integer topological invariants.\cite{Fidkowski11,Kitaev06,Supercohomology,FermionWW,Wang14,Metlitski14,Kapustin14} In the Floquet context, this reduction arises from a non-trivial interplay of the zero- and $\pi$- quasi-energy modes. In all cases, we find that the fermionic classification can be understood as having projective action of the symmetry group, $G$, (graded by fermion parity) combined with an effective integer valued time-translation symmetry under the Floquet evolution, leading us to hypothesize that such projective representations form a complete classification.

We then turn to the study of Floquet SPTs in bosonic systems (e.g. spin models). Here we build further evidence towards the hypothesized classification by constructing explicit models whose edge states realize all possible projective realizations of $G\times \Z$, where the extra factor of $\Z$ corresponds to discrete Floquet evolution ``symmetry". Interesting examples include, a dynamical analog of the Haldane spin-chain\cite{Haldane83,AKLT}, which exhibits free spin-1/2 edge states that flip under each driving period, and only come back to themselves after two periods. We also encounter bosonic examples where symmetries protect edge modes with quasi-energies that are neither $0$ nor $\pi$, but can be any rational fraction of $2\pi$.

Having built up a repertoire of concrete examples, we then formalize the hypothesized classification of 1D Floquet topological phases, by generalizing related classifications of equilibrium SPTs\cite{Chen11,Fidkowski11,Pollmann10} to periodically driven Floquet systems. We rigorously establish the above-hypothesized equivalence between the Floquet SPT classification with group $G$, and the equilibrium (``weak TI"-like) classification of $G\times \Z$ (or $G\rtimes \Z$ in the case of antiunitary symmetry group $G$). Recently, C. Von Keyserlingk and S. Sondhi presented a related but distinct classification with consistent results was demonstrated using a different method.\cite{Keyserlingk16}

Finally, we show that the requirement of many-body localization places strong requirements on the type of symmetry groups that can protect SPT phases. Specifically, via general representation theoretic arguments, we establish that symmetry-preserving many-body localized phases are impossible for symmetry groups with irreducible representations with dimension higher than one, which includes all non-Abelian unitary symmetries (e.g. spin-rotation), and also a various anti-unitary symmetries related to time-reversal. These results have far reaching consequences, beyond the context of Floquet systems, and for example rule out the possibility of many-body localization of electron topological insulators protected by time-reversal.

\section{Many-body localized (Floquet) Hamiltonians}
Since the requirement of many-body localization to avoid heating plays a crucial role in the sharp-distinction among interacting Floquet phases, we begin by reviewing a widely settled-upon sharp definition for the existence of many-body localizability. 

Full many-body localization is best defined through the existence of a complete set of quasi-local conserved quantities $\{n_\alpha\}$, that each take values $\{1\dots p_\alpha\}$, and together uniquely label an arbitrary eigenstate:
\begin{align}
|\Psi\>=|n_1 n_2\dots n_L\>
\end{align}

By quasi-local, it is meant that each $n_\alpha$ is exponentially well localized near a position, $r_\alpha$, i.e. that the projection operators:
\begin{align}
\Pi_{n_\alpha}=\sum_{n_{\beta\neq\alpha}}|n_1n_2\dots n_\alpha\dots n_L\>\<n_1n_2\dots n_\alpha\dots n_L|
\end{align}
differ from the identity at position $r$ by an exponentially small amount, i.e. for any local operator $\mathcal{O}(r)$ with bounded support near position $r$, $\frac{||[\Pi_{n_\alpha},\mathcal{O}(\v{r})||}{||\mathcal{O}(r)||}< e^{-|\v{r}-\v{r}_\alpha|/\xi}$, where $||\dots||$, and $\xi$ are an appropriate operator norm and localization length respectively. 

These projectors are exactly conserved quantities, that commute with the Hamiltonian, and hence their values are time-independent. More explicitly, the Hamiltonian of a static system can be written as a generic function of these projection operators $H_\text{MBL} = \sum_{\{n_\alpha\}} f\(\Pi_{n_1},\Pi_{n_2},\dots\)$, where $f$ is a (positive definite) quasi-local function of its arguments (i.e. is exponentially-weakly sensitive to the relative state of two-distant projectors). Or similarly, for a Floquet system, governed by a time-dependent Hamiltonian $H(t)$, which is periodic with period $T$, the time-evolution operator for a fixed period can be expressed as:
\begin{align}
F_\text{MBL} = \T e^{-i\int_0^T H(t) dt}=\prod_{\{n_\alpha\}} e^{if\(\Pi_{n_1},\Pi_{n_2},\dots\)}
\label{eq:FMBL}
\end{align}

In what follows, we will temporarily put aside the question of localization to focus on the topological aspects of Floquet phases. Our strategy will be to first construct examples of special zero-correlation length models that represent particularly simple realizations of various Floquet topological phases. After building some intuition from these simple models, we will give general arguments that the topological features of these zero-correlation length models are stable to generic perturbations and apply over a finite range of parameters, and in particular, will examine what constraints are placed by the requirement of localizability.

\section{Fermionic Floquet SPTs}
Having sharply defined a notion of many-body localizability, we now turn to the concrete task of systematically understanding 1D Floquet topological and SPT phases. We begin by reviewing some previously known constructions of topological phases in non-interacting fermionic systems with periodic driving, and then address how these results are modified upon the inclusion of interactions.

\subsection{Floquet Majorana modes in non-interacting models \label{sec:Noninteracting}}
 In a non-interacting static superconducting wire, BdG excitations with energy $E$ and $-E$ are related by particle-hole conjugation, and correspond to complex fermionic excitations unless $E=0$. However, in a driven system, quasi energy $\e$ is defined modulo $2\pi$ (here, and throughout, we normalize the quasi-energy with respect to the Floquet period $T$, such that quasi-energies become dimensionless phases between $0$ and $2\pi$), and hence $\e =\pm \pi$ are equivalent enabling real (i.e. self-conjugate) Majorana modes at energy $\pi$ . To set the stage for the study , we describe a simple toy model\cite{Asboth14,Gannot15,Khemani15} that exhibits perfectly localized Floquet Majorana Pi Modes (MPMs), i.e. real (self-conjugate) fermionic modes with quasi-energy exactly quantized to $\pi$ localized to the edge of a driven superconducting wire. Previous works have given more experimentally achievable proposals for realizing these phases\cite{Jiang11}, however the toy models will be instructive for establishing the proof of existence for more general Floquet SPTs, and analyzing the effects of interactions.
 
\begin{figure}[tb]
\begin{center}
\includegraphics[width=0.8\linewidth]{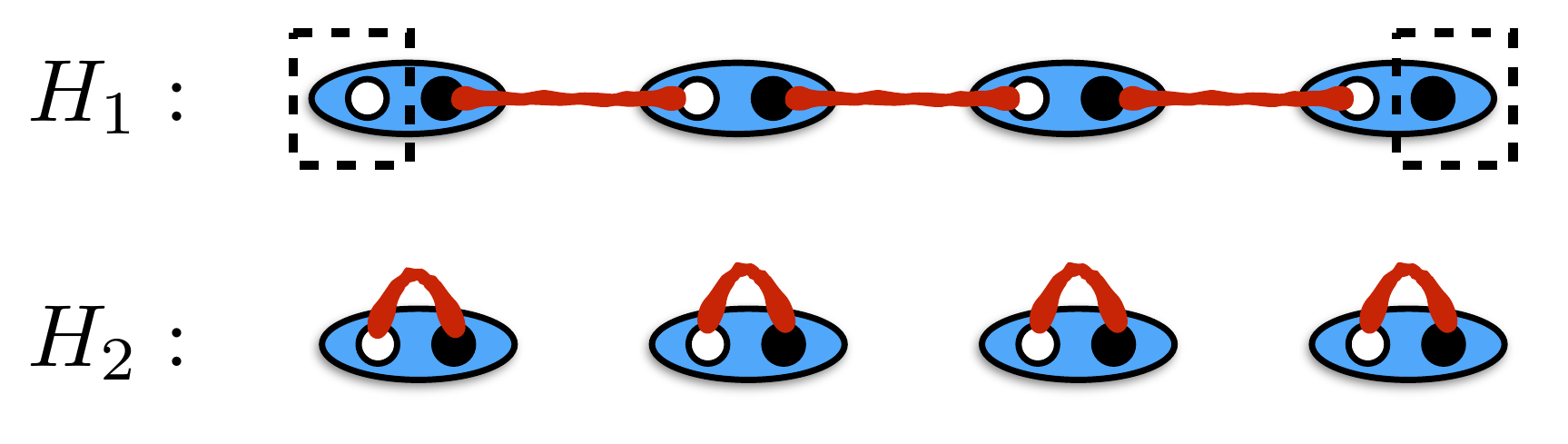}
\end{center}
\caption{
{\bf Schematic picture of stroboscopic Floquet drive.} The phase with Floquet Majorana edge states is obtained by alternating time-evolution with a topological Hamiltonian $H_1$, for time $T_1$ followed by time-evolution under a trivial Hamiltonian, $H_2$, for time $T_2$. Complex fermions $c_j$ (blue ovals) are decomposed into two Majorana fermions $a_j,b_j$ (white and black circles respectively). Non-zero couplings are represented by wavy red segments.
}
\label{fig:Floquet}
\end{figure}

Consider a superconducting chain of spinless (complex) fermions $c_j = \frac{1}{2}\(a_j+ib_j\)$, where $j$ labels sites of the chain, and $a$ and $b$ are real (Majorana) fermion operators satisfying cannonical anti-commutation relations: $\{a_i,a_j\} = 2\delta_{ij} = \{b_i,b_j\}$, $\{a_i,b_j\} =0$. A particularly simple construction that realizes a nontrivial Floquet topological phase is obtained by subjecting the chain to a ``stroboscopic" periodic time-dependent drive under Hamiltonian:
\begin{align}
H(t) = 
	\begin{cases}
	H_1  = \frac{i\lambda_1}{4}\sum_{j=1}^L a_jb_j & 0\leq t < T_1 \\[6pt]
	H_2  = \frac{i\lambda_2}{4}\sum_{j=1}^{L-1}b_j a_{j+1}& T_1\leq t < T_1+T_2
	\end{cases}
	\label{eq:HMPM}
\end{align}
For this alternating drive, the time-evolution over the duration of a single period (Floquet operator), $T=T_1+T_2$, decomposes into the product $F = \T e^{-i\int_0^{T}H_0(t)dt}=F_2F_1$ with $F_j = e^{-iH_jT_j}$. Here $H_1$ and $H_2$ are respectively zero-correlation length ``fixed-point" Hamiltonians for the trivial and topological phases of the superconducting chain. In particular, $a_0$ and $b_L$ do not appear in $H_2$, and hence would be local Majorana zero quasi-energy modes (MZMs) for $T_1=0$. 

If we instead choose $\frac{\lambda_1T_1}{4} = \frac{\pi}{2}$, which, using the identity, $e^{\theta ab} = \cos\theta+\sin\theta ab$, gives: $F_1 = \prod_{j=1}^L a_j b_j = a_1 \(\prod_{j=1}^{L-1}b_j a_{j+1}\) b_L = a_1 e^{i 2\pi H_2/\lambda_1} b_L$, the full Floquet time-evolution operator reads:
\begin{align}
F = a_1 e^{-i\tilde{T}_2 H_2} b_L \equiv e^{-iH_F}
\end{align}
where $\tilde{T}_2 = T_2 - 2\pi /\lambda_1$, and $H_F = \tilde{T}_2H_2+\frac{i\pi}{2}a_1b_L$ is the Floquet Hamiltonian for a specific branch cut of $\log F$.

We note that $a_1$ and $b_L$ are left out of $H_2$, and hence commute with $H_2$. Then $Fa_1F^\dagger = b_La_1b_L = -a_1$ and similarly $Fb_LF^\dagger = -b_L$. Hence, $a_1$ and $b_L$ are localized Majorana fermion modes with $\pi$ quasi-energy, which we will henceforth refer to as Majorana-Pi-Modes (MPMs).

While we have so far demonstrated the existence of the strictly localized MPMs only for a particular choice of parameters, the MPMs are stable against small perturbations of the driving Hamiltonian and persist over a finite range of parameters centered around the ones chosen above. Just as in non-driven equilibrium quantum systems, to assess the stability of the MPMs to generic small local perturbations of the Hamiltonian $H(t)\rightarrow H(t)+V(t)$ by focusing on allowed local interactions involving the topological edge modes, and ignoring bulk degrees of freedom. Specifically, a generic local time dependent perturbation $V(t)$, induces a quasi-local change in the Floquet Hamiltonian: $F = e^{-iH_F}\rightarrow F' = e^{-i\(H_F+\Delta H_F\)}$. Due to Lieb-Robinson bounds on the dynamical spread of the influence of a local perturbation, since $V$ is local, $\Delta H$ will also be quasi-local, i.e. consists of exponentially well localized terms. For small $V$, i.e. with operator norm $|V(t)|\ll \frac{1}{T}$, the explicit form of $\Delta H_F$ for a given $V(t)$ may be computed through standard time-dependent perturbation theory. 

However, for our purposes it is instead sufficient to consider stability of the edge modes against the addition of a generic quasi-local perturbation, $\Delta H_F$, in the Floquet Hamiltonian. For $\tilde{T_2}\lambda_2\neq 2\pi$, the bulk degrees of freedom have quasi-energy different from $\pi$, and are hence separated by an energy gap from mixing with the MPMs at the ends of the wire. Consequently, just as for topological zero modes in static systems, sufficiently small perturbations that mix the MPMs with bulk degrees of freedom simply virtually dress the MPMs with an amplitude decaying exponentially with characteristic distance $\xi\lesssim \(\log\frac{\tilde{T}_2\lambda_2}{|V|T}\)^{-1}$ away from the edge of the wire. Hence, due to the locality of the perturbation $\Delta H_F$, the change in the coefficient of the non-local term $\frac{i\pi}{2}a_0b_L$ will be exponentially small in $e^{-L/\xi}$, such that the quasi-energy of these modes is topologically protected at $\pi$ for asymptotically long wires ($L\rightarrow \infty$).

The above-described model with MPMs serves as a basic building block for constructing general fermionic Floquet SPTs. To this end, we may consider $N_0$ chains of fermions $c_n = \frac{1}{2}\(a_n+ib_n\)$, with flavor index $n = 1\dots N_0$, driven by Eq.~ \ref{eq:HMPM} with $\lambda_1=0$, and $N_\pi$ chains of fermions $\psi_m = \frac{1}{2}\(\alpha_m+i\beta_m\)$ driven by Eq.~\ref{eq:HMPM} with $\lambda_1 = 2\pi/T_1$. These respectively result in $N_0$ MZMs, $\(a_{n,1},b_{L,1}\)$ and $N_\pi$ MPMs, $\(\alpha_{m,1},\beta_{m,1}\)$ that are strictly localized to the ends of the chain. 

\subsection{No Symmetry}
In the absence of symmetry, there is no topological protection for an even number of MZMs or MPMs. To see this, we may restrict our attention to possible perturbations within the Hilbert space spanned by the topological modes at one end of the chain, since, coupling the Majorana end-states to complex bulk degrees of freedom will simply renormalize the spatial extent of their wave-function without perturbing their quasi-energy. For concreteness, consider the left end of the chain. The most general non-interacting coupling terms involving the topological modes $\{a_{n,1},\alpha_{m,1}\}$ that can be generated by a T-periodic perturbation to $H(t)$ is: $\Delta H_F = \frac{i}{4}\sum_{n\neq n'} a_{n,1}M^{(0)}_{n,n'}a_{n',1}+\frac{i}{4}\sum_{m\neq m'} \alpha_{m,1} M^{(\pi)}_{m,m'}\alpha_{m',1}$, where $M^{(0,\pi)}$ are antisymmetric matrices. 

\subsubsection{Dynamical decoupling of $0$ and $\pi$ modes}
Note that bilinear couplings between MZMs and MPMs are ineffective, and can be ignored. As shown in Appendix~\ref{app:noninteracting}, such couplings may be eliminated by defining new MZM and MPM operators from linear combinations of $a_{n,0},\alpha_{n,0}$. 

\begin{figure}[tb]
\begin{center}
\includegraphics[width=0.5\linewidth]{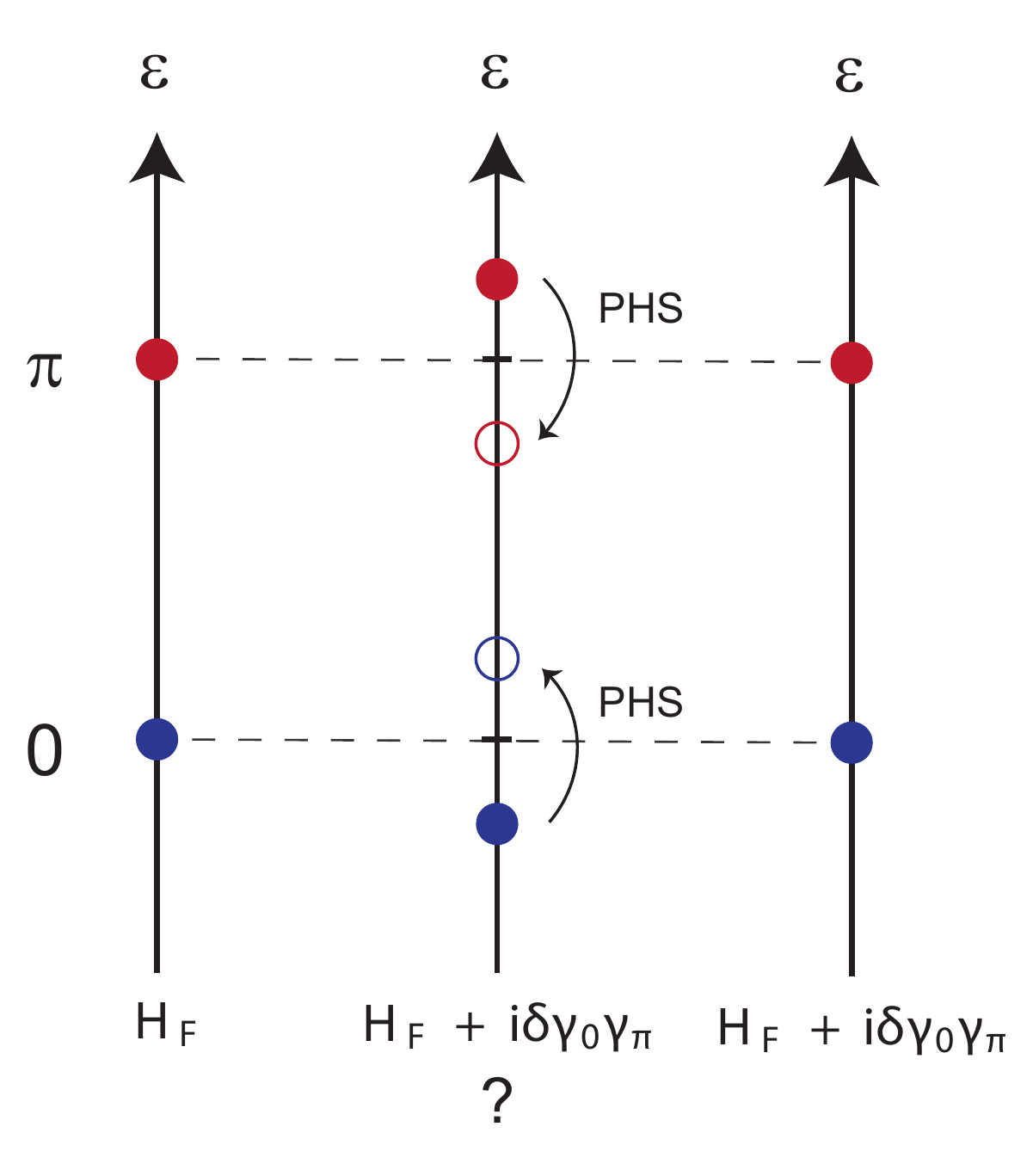}
\end{center}
\vspace{-0.2in}
\caption{
{\bf Schematic picture of quasi-energy spectrum of non-interacting fermionic Floquet SPTs.} The particle-hole symmetry of BdG equations indicates that a bilinear coupling between a MZM and a MPM is ineffective, and cannot move the MZM or MPM away from $0$ or $\pi$ quasi-energy respectively. The leftmost line shows the unperturbed quasi-energies. The middle line illustrates that any possible splitting due to the perturbation does not respect particle-hole symmetry (PHS), indicating that the resulting quasi-energies (rightmost line) must be identical to the initial unperturbed quasi-energies $0$ and $\pi$ (leftmost line)
}
\label{fig: noninteracting}
\end{figure}

A more general argument establishing that MZM and MPMs cannot split each other via non-interacting couplings can be obtained by considering the effective particle-hole ``symmetry" of BdG Hamiltonians, which dictates that single particle levels with quasi energy $\(\e\mod 2\pi\)$ must be related by particle-hole conjugated levels with quasi-energy $\(-\e\mod2\pi\)$. For self-conjugate (real) Majorana modes, this requires that $\(\e\mod 2\pi\) = \(-\e\mod 2\pi\)$, which has only two discrete solutions $\e=0,\pi$ -- the latter solution is only possible in a periodically driven system where energy is only conserved modulo $2\pi$, highlighting the special features of the Floquet-driven system. Naively, turning on a weak non-interacting coupling, $i\delta\gamma_0\gamma_\pi$ of strength $\delta\ll 1$ between a MZM $\gamma_0$ with quasi-energy $\e_1=0$, and a MPM $\gamma_\pi$ with quasi-energy $\e_2=\pi$, would split their quasi-energies into $\e_1 = -\mathcal{O}(\delta)$, and $\e_2 = \pi +\mathcal{O}(\delta)$. However, as illustrated in Fig.~\ref{fig: noninteracting}, one can easily see that this outcome is not compatible with particle-hole symmetry (i.e. in this scenario there would be no particle-hole conjugate modes at $\tilde{\e}_1 = +\mathcal{O}(\delta)$, and $\tilde{\e}_2 = \pi - \mathcal{O}(\delta)$). Hence, the only possible outcome of turning on the strength-$\delta$ coupling, is that the new eigen-modes continue to have quasi-energy $\e_1 = 0$ and $\e_2 = \pi$ respectively. In a loose sense, such couplings can be thought of as ``forbidden", since they do not conserve quasi-energy modulo $2\pi$ (as defined in terms of the unperturbed Hamiltonian).

\subsubsection{Classification of Floquet phases in the absence of symmetry}
For $N_{0}$ ($N_\pi$) odd, $H_F+\Delta H_F$ inevitably exhibits a single unpaired Majorana mode with $0$ ($\pi$) quasi-energy. However, for $N_0$ ($N_\pi$) even $\Delta H_F$ can pair the Majorana end-states into complex fermions and split them away from $0$ ($\pi$) quasi-energy. Hence, we see that there are 4 distinct Floquet topological states in the absence of symmetry, characterized by a pair of $\Z_2$ invariants indicating the parity of $N_0$ and $N_\pi$, corresponding to a $\Z_2\times \Z_2$ classification of phases. 

For any of the topologically non-trivial Floquet phases, there are either $1$ or $2$ edge-modes, and hence turning on 4-fermion or higher interaction terms does not generate any new possible couplings among the edge states. Hence, we expect the non-interacting classification in the absence of symmetry to coincide with the interacting one. One the other hand, certain symmetries can protect larger numbers of edge modes, in which case interactions offer additional ways to gap out the non-interacting topological edge modes and alter the SPT classification.\cite{Fidkowski11}

\subsection{Interacting fermionic Floquet SPTs}
The presence of a global symmetry group, $G$, constrains the possible form of perturbations (i.e. restricts the entries of $M^{(0,\pi)}$), and can protect multiple MZM and MPMs. Since non-interacting terms cannot mix the MZMs and MPMs, the analysis of symmetry-allowed mass terms $M^{(0)}$ and $M^{(\pi)}$ each independently follow exactly from the analysis for static non-interacting SPT phases.  which are well understood, and for a given group $G$ the group of distinct fermionic SPT phases arising from non-interacting static Hamiltonians, $C^{(\text{NI})}_\text{st}[G]$, is known.\cite{Kitagawa10,Jiang11,Rudner13,Asboth14,Gannot15}  In all cases, non-trivial static 1D SPT phases are characterized self-conjugate zero-energy edge states. In the non-interacting Floquet context, the most general new possible phases arise from the possibility of also realizing self-conjugate modes at quasi-energy $\pi$. Hence, from the above considerations, we see that the non-interacting classification of periodically driven Floquet SPT phases then simply yields two independent copies of the non-interacting band-invariants -- one each for $0$ and $\pi$ quasi-energy modes., corresponding to a non-interacting Floquet classification: 
\begin{align}
C^{(\text{NI})}_F[G]=C^{(\text{NI})}_\text{st}[G]\times C^{(\text{NI})}_\text{st}[G]
\end{align}

For static SPTs, interactions can modify the free-fermion classification. Specifically, in many cases where $C_\text{st}^{(\text{NI})} = \Z$, the interacting classification is reduced to $C_\text{st}^{(\text{NI})}\rightarrow C_\text{st}=Z_N$ where $N$ is some even integer. A simple guess based on the above considerations would be that the corresponding Floquet classification would again follow simply from the static classification as: $C_F \overset{?}{=} C_\text{st}\times C_\text{st}$. However, we will see that the situation is more subtle, and that interactions can effectively enable Floquet analogs of Umklapp type terms that conserve quasi-energy only modulo $2\pi$ that can mix the MPM and MZM sectors in non-trivial ways.

To understand the whether interactions reduce the non-interacting classification, we again consider perturbations, $\Delta H_F$, to the Floquet Hamiltonian that couple the topological zero- and $\pi$- quasi-energy modes, but allow for interaction terms involving two- or higher- body interaction terms involving products of $4$ or more edge modes. 

For concreteness, we start with the specific illustrative example of spinless, time-reversal (TR) symmetric superconducting chains, corresponding to Altland-Zirnbauer (AZ) class BDI, and then give general results for all of the symmetry classes corresponding to the 10-fold way.

\subsection{Spinless TR-invariant superconductors (class BDI)} 
Before diving into the analysis of the interacting Floquet phases, a comment on the notion of time-reversal (TR) symmetry in time-dependent quantum systems is in order. Whereas for static Hamiltonians, the dynamics may be invariant in the reversal of time about any reference time, $t_0$, for periodic time-dependent Hamiltonians, can at most exhibit a discrete set of time inversion centers, $t_0$ such that: $H(t_0+t) = \T H(t_0-t)\T^{-1}$. Acting on the Floquet Hamiltonian, $H_F = i\log F$,  such discrete time-reversal of the drive acts as an ordinary anti-unitary operator $\T$, just as for a static Hamiltonian. In particular, the Floquet evolution for the full period transforms under time-reversal as: $\T F\T^{-1} = \T e^{-iH_F} \T^{-1} = e^{+i\T H_F \T^{-1}}$. For time-reversal invariant Floquet Hamiltonians, $\T H_F\T^{-1} = H_F$, this implies that:
\begin{align}
\T F \T^{-1} \overset{(\text{TRS})}{=} F^{-1}
\end{align}

For example, the Hamiltonian, Eq.~\ref{eq:HMPM}, exhibits a time-reversal symmetry that defined by: $\T c_n\T^{-1} = c_n$ $\T\psi_m\T^{-1}=\psi_m$, i.e.:
\begin{align}
\T\begin{pmatrix} a_n \\ b_n \\ \alpha_m \\ \beta_m \end{pmatrix}\T^{-1}
=
\begin{pmatrix} a_n \\ -b_n \\ \alpha_m \\ -\beta_m \end{pmatrix}
\label{eq:TRBDI}
\end{align}
The relative minus signs in Eq.~\ref{eq:TRBDI} indicate that single particle couplings between MZM or MPM edge states of the form $ia_na_{n'}$ or $i\alpha_m\alpha_{m'}$ are odd under TR and hence forbidden by symmetry. Moreover, as previously remarked, time-periodic single particle couplings cannot mix zero and $\pi$ quasi-energy modes. Hence, the non-interacting phases are characterized by two integer invariants $\(N_0,N_\pi\)$ respectively indicating the number of MZMs and MPMs localized to a boundary of the wire, corresponding to a $\Z\times\Z$ classification of Floquet SPT phases. 

We note that, fermionic Hamiltonians may only include terms with even numbers of fermion operators, and are therefore inevitably invariant under the fermion parity operator $P_f=(-1)^{N_F}$, where $N_F$ is the total number of fermions in the system. For this reason, fermion parity is sometimes taken to be part of the symmetry group of a fermionic system. Since $P_f^2 = 1$, this corresponds to an extra factor of $\Z_2$ in $G$, typically denoted $\Z_2^F$. However, unlike a conventional $\Z_2$ symmetry, $\Z_2^F$ cannot be broken even spontaneously by interactions.

\renewcommand{\arraystretch}{1.5}
\begin{table}
\begin{tabular}{C{1.1in}C{2.1in}}
\toprule
{\bf Phase} $(N_0,N_\pi)$ &  {\bf Defining Edge Characteristic}  \\ 
\midrule
(1, 0) & unpaired Majorana \\
(2, 0) & $\T P_f =- P_f \T $ 
\\
~~~~~~(4, 0)~~(B) &  $\T^2=-1$\\
(1,-1) & $FP_f = -P_fF$ \\
~~~~~~(2,-2)~~(B)  & $\T F\T^{-1} = -F $ \\
\bottomrule
\end{tabular}
\caption{{\bf Time-reversal symmetric 1D fermion FSPTs -} The classification of 1D interacting fermionic Floquet SPTs with time reversal (class BDI) has a $\Z_8\times \Z_4$ group structure. Topological phases are labeled by the number of Majorana zero and $\pi$ quasi-energy modes, $\(N_0,N_\pi\)$, present in the absence of interactions. The projective edge algebra that defines these phases are shown in the right column. The phases labeled (B) are topologically equivalent to bosonic FSPT phases. }
\end{table}

While no single-particle perturbations can disturb these topological edge modes, it is known from the study of static systems that four-body interactions can fully remove the degeneracy associated with eight MZMs (or similarly, with any integer multiple of eight MZMs)\cite{Fidkowski11,TurnerPollmannBerg}. We briefly recall the key ideas behind this result. First, note that any arbitrary interaction involving an odd number of MZMs will leave behind at least one exact MZM, such that any phase with odd $N_0$ is nontrivial. 

For $N_0=2$, the most general edge-state perturbation is the non-interacting term $P_f^{\text{(loc)}}=ia_1a_2$, which is odd under time-reversal symmetry and hence cannot be generated by any symmetry preserving perturbation. The operator $P_f^{\text{(loc)}} = 2f^\dagger_{12}f^{\vphantom\dagger}_{12}-1$ squares to $1$, and consequently has eigenvalues $\pm 1$, corresponding to the fermion parity of the complex fermion zero mode $f_{12} = \frac{1}{2}\(a_1+ia_2\)$. Therefore, for $N_0 = 2$, $P_f^{\text{(loc)}}$ represent the local action of the fermion parity operator acting within the low-energy subspace spanned by MZM edge states. While, the $\T$ commutes with the total fermion parity, $P_f$, it anticommutes (i.e. commutes only up to an overall phase of $-1$) with the local action of fermion parity on the edge state zero modes: $\T P_f^{\text{(loc)}}\T^{-1} = -P_f^{\text{(loc)}}$, and thus the $N_0=2$ MZM edge forms a projective representation of $\Z_2^T\times\Z_2^F$. It is generally true that static 1D SPT phases are systematically classified by projective representation of $G$ (for bosonic systems) or $G\times\Z_2^F$ for (fermionic systems). Namely, any non-trivial projective action of symmetry action on the edge modes requires an edge-mode Hilbert space of dimension larger than one (all 1D representations of $G$ are Abelian and hence non-projective) -- i.e. requires an edge state degeneracy that cannot be lifted without sacrificing symmetry. Moreover, since local bulk degrees of freedom necessarily transform under an ordinary representation of the symmetry group, there is no way for them to form a non-degenerate symmetry singlet by interacting with the edge modes. The projective representations form an Abelian group, with each group element corresponding to a distinct static topological phase of matter.

For $N_0=4$ MZMs $a_{1\dots 4}$, the interaction term $V = \lambda\(ia_1a_2\)\(ia_3a_4\)$ is allowed by symmetry. This divides the 4-fold degenerate space spanned by the 4 MZMs into two doublets: $\{|00\>,|11\> \equiv f_{34}^\dagger f_{12}^\dagger|00\>\}$, and $\{|10\> \equiv f_{12}^\dagger|00\>,|01\> \equiv f_{34}^\dagger|00\>\}$ labeled by the occupation numbers of $f_{12} = \frac{1}{2}\(a_1+ia_2\)$ and $f_{34} = \frac{1}{2}\(a_3+ia_4\)$. However, the smaller two-fold degeneracy of these doublets is protected by symmetry and cannot be removed by any symmetry preserving perturbation. To see this, consider the subspace spanned by one such doublet, say, $\{|00\>,|11\>\}$, and define Pauli-like spin operators $\sigma^z = |00\>\<00|-|11\>\<11|$, $\sigma^x = |00\>\<11|+|11\>\<00|$, and $\sigma^y = -i\(|00\>\<11|-|11\>\<00|\)$. Since both $|00\>$ and $|11\>$ have even fermion parity, $P_f$ acts like the identity operator in the subspace of this even doublet. However, the action of time-reversal on the doublet is unconventional. Namely, note that $\T f_{ij}^{\vphantom\dagger} \T = f_{ij}^\dagger$, and hence time-reversal flips the occupation number of the two zero modes. Hence we may choose the relative phase of $|00\>$ and $|11\>$ such that $\T|00\> = |11\>$. On the other hand, $\T^2|00\> = \T|11\> = \T f_{12}^\dagger \T^{-1}\T f_{34}^\dagger\T^{-1}\T|00\> = f^{\vphantom\dagger}_{12}f^{\vphantom\dagger}_{34}f_{12}^\dagger f_{34}^\dagger|00\>= -|00\>$. This is another example of projective action of symmetry, since $\T^2=+1$ on any bulk degree of freedom, whereas $\T^2=-1$ on the edge mode doublets -- indicating that the edge modes form a Kramers doublet whose two-fold degeneracy is protected by $\T$. We note that the fact that fermion parity acts trivially on the local edge states indicates that the presence of local fermion degrees of freedom was unimportant for realizing this particular SPT order. Indeed, the same projective realization of symmetry can be realized by the edge modes of a purely bosonic SPT, indicating that the fermionic system with $N_0=4$ reduces to a bosonic one in the presence of interactions.

For $N_0 = 6$, the edge modes transform as a combination of the projective properties of $N_0=2$ and $N_0=4$, namely, the local action of symmetry on the edge states satisfies $\T P_f\T^{-1} = -P_f$ and $\T^2=-1$, indicating a protected 4-fold symmetry.

For $(N_0,N_\pi) = (8,0)$, we can readily see that doubling the $(-1)$ phase factors for the above described $(N_0,N_\pi) = (4,0)$ results in an ordinary (non-projective) action of symmetry on the MZM edge states, indicating that there is no special topological protection of these modes. Indeed, we can concretely confirm this suspicion by combining the 8 MZMs into two bosonic doublets, one consisting of the even fermion parity configurations of $a_{1\dots 4}$ and another from those of $a_{5\dots 8}$. In analogy to the $N_0=4$ case described above, can introduce the Pauli operators $\sigma_1$ and $\sigma_2$ acting on each of these doublets, which both transform like Kramers doublets under $\T$. However, e.g. the Heisenberg interaction $V = \boldsymbol{\sigma}_1\cdot\boldsymbol{\sigma}_2$ clearly preserves $\T$, despite the Kramers nature of $\sigma_{1,2}$, and removes the degeneracy of the zero modes by selecting a pseudo-spin singlet combination of $\sigma_{1,2}$.

We now perform a similar analysis of the perturbative stability for edge modes of the periodically driven system. To conserve quasi-energy modulo $2\pi$, interacting edge perturbations must involve Floquet-Umklapp type terms that couple even numbers of MZMs and MPMs. Hence, phases in which $N_0-N_\pi$ is odd remain non-trivial, in particular even when the total number of topological edge modes $N_0+N_\pi$ is an integer multiple of $8$. Again, by repeating the above considerations from static systems, one can easily verify that $4$ MPMs modes can be symmetrically coupled to produce a degenerate bosonic doublet spanned by the spin-1/2 operators $\boldsymbol{\sigma_\pi}$, which transforms as $\T^2=-1$ under time-reversal, and is static under the Floquet evolution, just as for the non-driven phase with four MZMs. 

Hence, for a phase with $(N_0,N_\pi) = (4,-4)\simeq (4,4)$, we may add the symmetry preserving interaction $-V\boldsymbol{\sigma}_0\cdot \boldsymbol{\sigma}_\pi$ to completely lift the edge degeneracy. This shows that the non-interacting $(4,-4)$ phase reduces to a trivial phase in the presence of interactions, due to the non-trivial Floquet-Umklapp interaction between MZMs and MPMs, i.e. that having 4 MZMs is topologically equivalent to having 4 MPMs.

On the other hand the $(2,-2)$ state remains topologically non-trivial even in the presence of interactions due to a dynamical winding property. To see this, let us start with the non-interacting $(2,-2)$ state and, as before, add an edge perturbation $\sim a_1a_2\beta_1\beta_2$, to break the edge state sector into a bosonic  degrees of freedom. For example, in the even-fermion parity sector with $|00\>,|11\> = f^\dagger \psi^\dagger|00\>$, where $f = \frac{1}{2}\(a_1+ia_2\)$ is a complex zero mode, and $\psi = \frac{1}{2}\(\beta_1+i\beta_2\)$ is a complex $\pi$ mode, we may define the bosonic pseudospin: $\sigma^z = |00\>\<00|-|11\>\<11|$. Since $f$ and $\psi$ are conjugated by $\T$, $\T$ must flip the state of $\sigma^z$. However, since these complex fermions acquire a relative $(-1)$ phase under $\T$: $\T f\T^{-1} = f^\dagger$, $\T\psi\T^{-1}=-\psi$, $\sigma$ behaves like a non-Kramers singlet ($\T^2=1$), under $\T$. Hence, we may represent the local action of $\T$ on the $(2,-2)$ edge as $\T = \sigma^xK$. On the other hand, $|00\>$ and $|11\>$ have quasi-energies that differ by $\pi$, and hence acquire a relative $(-1)$ phase under the Floquet evolution, such that the local action of Floquet time-translation on the edge states is represented as: $F = \sigma^z$. Combining these two properties, we see that $\T$ and $F$ act projectively on the topological edge states of the $(2,-2)$ phase:
\begin{align}
\T F \T^{-1}=(-1)F^{-1}
\end{align}
in contrast to the non-projective action ($\T F\T^{-1} = F^{-1}$) for bulk degrees of freedom. This non-trivial projective edge action holds, even though the static symmetry group generated by $\T,P_f$ acts trivially on the edge. We can picture this phase as having a free psuedo-spin-1/2 edge degree of freedom, $\sigma$, that rotates by $\pi$ around the z-axis over the coarse of each period. While, we could add a symmetry-preserving field $h\sigma^x$ to try to pin this edge spin, however the effect of this field would average to zero over a sequence of two driving periods due to the non-trivial Floquet dynamics of the edge spin. 

The set of interaction floquet SPT phases that arises from these considerations can be generated by combinations of two "root" phases: $(N_0,N_\pi)=(1,0)$ and $(1,-1)$. N-fold combinations of the former phase for $N=0,1,\dots 7$ realize all of the static, non-driven topological phases that realize projective edge-representations of time-reversal (and fermion parity). The latter sequence of phases generated by combinations of, $(1,-1)$, transforms ordinarily under the static symmetry group, but has non-trivial interplay of symmetry and topological Floquet dynamics that produce a projective edge-action of symmetry and time-translation. The $(1,-1)$ phase has $\{F,P_f\}=0$ at the edge, the $(2,-2)$ phase has $\{F,\T\}=0$, the $(3,-3)$ phase has both $\{F,P_f\}=0$ and $\T F\T^{-1}F=-1$, and the $(4,-4)$ phase is trivial and should be identified with $(0,0)$. 

\begin{figure}[tb]
\begin{center}
\includegraphics[width=0.9\linewidth]{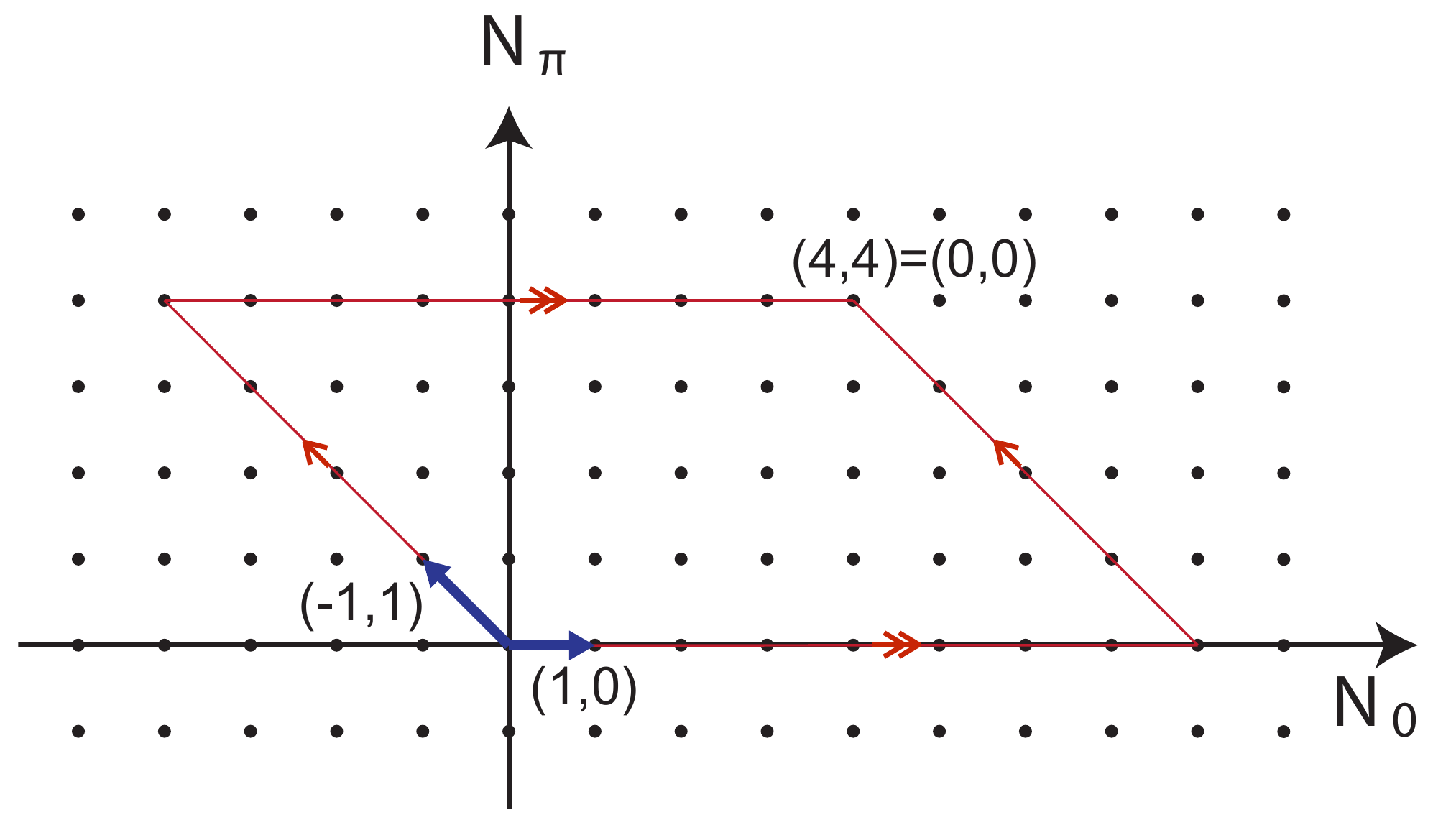}
\end{center}
\caption{
{\bf Group structure of interacting Floquet SPT phases of spinless TR-invariant superconductors (class BDI).} In the absence of interactions, each point corresponds to a topological phase with $N_0$ MZMs and $N_\pi$ MPMs at the edge. With interactions, only points on a discrete torus (bounded red region, with arrows indicating periodic boundary conditions) correspond to distinct topological phases of the $\mathbb{Z}_8 \times \mathbb{Z}_4$ classification. The blue vectors $(1,0)$ and $(-1,1)$ are generators of the subgroups $\mathbb{Z}_8$ and $\mathbb{Z}_4$, respectively.
}
\label{fig: Z8xZ4}
\end{figure}

Thus we see that the non-interacting Floquet classification has been reduced from $\Z\times \Z$ to $\Z_8\times \Z_4$. The associated group structure of the SPT phases are shown in Fig.~\ref{fig: Z8xZ4}. These phases exhaust all possible projective representations of $\T,P_f,$ and $F$, which leads us to hypothesize that the full classification of interacting Floquet SPT phases with symmetry group $G$ is given by the group of projective representations of $G\times \Z$ graded by $Z_2^{F}$ fermion parity ``symmetry", where the extra factor of $\Z$ corresponds to time-translation symmetry.

\subsection{Other symmetry groups}
We have so far analyzed the case of no symmetry (class D) and spinless time-reversal symmetry (class AIII). We can repeat the above perturbative stability analysis of the non-interacting Floquet classifications for the other non-trivial 1D SPT symmetry classes in the 10-fold way. At a first pass, we will ignore the requirement of many-body localizability, necessary to avoid runaway heating by the drive frequency, and just study the topological outcomes. It will turn out that only the cases with no symmetry (class D) and spinless time-reversal (BDI) permit a many-body localized Floquet SPT phases that are stable against heating. The other non-trivial 1D symmetry classes all have group structures that have irreducible representations of dimension greater than one, which we will show below, protect local degeneracies that spoil the possibility of having a symmetry-preserving many-body localized phase. However, it is instructive to consider other examples to build our intuition. Moreover, if interactions are weak, it is conceivable that the Floquet SPT phases described in these other classes may survive without catastrophic heating for adequately long times to be of interest for experiments.

The results are summarized in Table. \ref{tab:FermionPhases}. In each case, we find precise agreement between the stability analysis and projective representations of $G\times \Z$ graded by $\Z_2^F$ fermion parity ``symmetry", further supporting the hypothesis that this represents a complete classification. 

For Kramers-doublet fermions with time-reversal symmetry (class DIII), the non-interacting classification is unchanged by interactions. The non-trivial phases are characterized by two $\Z_2$ invariants that represent the presence or absence of a Kramers pair of MZMs or MPMs respectively.

Systems with fermions with a conserved $U(1)$ charge that is time-reversal odd, $G = U(1)\times\Z_2^T$, (class AIII), derive directly from the study with only time-reversal (BDI), but only the sub-group of phases with an even number of $N_0,N_\pi$ are compatible with the $U(1)$ charge conservation. Equilibrium examples of this class include spinless fermions with random hopping amplitudes, which satisfy in an anti-unitary particle-hole symmetry that may be regarded as ``time-reversal". 

Finally, in the absence of interactions systems with a conserved $U(1)$ charge that and both discrete time-reversal and charge conjugation symmetries, $G = U(1)\rtimes\(Z_2^T\times\Z_2^C\)$, realize only the subset of BDI phases with multiples of $4$ MZMs or MPMs. Since $4$ MPMs are equivalent to $4$ MZMs, in the presence of interactions, the periodic driving does not enable any new non-equilibrium phases.

\renewcommand{\arraystretch}{1.5}
\begin{table}
\begin{tabular}{C{1.2in}C{0.4in}C{0.7in}C{1in}}
\toprule
{\bf Symmetry} & {\bf AZ Class} & {\bf Static (free$\rightarrow$int.)} & {\bf Floquet (free$\rightarrow$int.)} \\ 
\midrule
{\centering None} & D & $\Z_2$ & $\Z_2\times\Z_2$ \\ 
$\Z_2^T$,~\scalebox{0.7}{$(\T^2=P_f)$} & DIII & $\Z_2$ & $\textcolor{red}{\Z_2\times \Z_2}$ \\ 
$\Z_2^T$ \scalebox{0.7}{$(\T^2=1)$} & BDI & $\Z\rightarrow\Z_8$ & $\Z\times\Z\rightarrow\Z_8\times\Z_4$ \\ 
$U(1)\times \Z_2^T$ & AIII & $\Z\rightarrow\Z_4$ & $\Z\times\Z\rightarrow\textcolor{red}{\Z_4\times\Z_2}$ \\ 
$U(1)\rtimes  \(Z_2^T\times\Z_2^C\)$, \scalebox{0.7}{$(\T^2=P_f)$} & CII & $\Z\rightarrow\Z_2$ & $\Z\times\Z\rightarrow\textcolor{red}{\Z_{2}}$ \\
\bottomrule
\end{tabular}
\caption{{\bf Classification of 1D fermionic Floquet SPTs -- } Group structure of nontrivial topological classes for 1D fermionic systems with discrete on-site symmetries, listed by physical symmetry group and equivalent Altland-Zirnbauer (AZ) class. $C\rightarrow C'$ indicates that the non-interacting classification $C$ is changed by interactions to $C'$. Red text indicates symmetry groups that are incompatible with many-body localization, and are therefore unstable to runaway heating in the presence of generic bulk interactions.
\label{tab:FermionPhases}}
\end{table}

\section{Bosonic Floquet SPTs}
In the above fermionic classification of Floquet SPTs with symmetry groups in the 10-fold way with some combination of charge-conservation, time-reversal and particle-hole symmetries, we found that only topological superconducting classes (i.e. those without a conserved $U(1)$ charge) permit stable, localizable Floquet SPTs. Unfortunately, in quantum optics based setups, such as cold-atoms, superfluid phases are unsuitable to many-body localization due to the presence of a non-localizable Goldstone mode. 

To uncover potentially experimentally relevant Floquet SPT phases in fermionic systems, we would need to look into other symmetry groups. Alternatively, we may examine the prospect of finding Floquet SPTs in bosonic (e.g. spin) systems. Here, we will many-examples of localizable Floquet SPTs that are stable against heating.

\subsection{Time-reversal symmetry $(\T^2=1)$}
To begin, let us consider bosonic models, such as integer-spin chains, with time-reversal symmetry, that squares to unity. The ground-state classification of such systems includes a single non-trivial phase first explained by Haldane\cite{Haldane83,AKLT}, which exhibits free spin-1/2 edge states that transform as $\T_\text{edge}^2=-1$ under time-reversal symmetry. This Haldane phase is not localizable with full spin-rotation symmetry, however, one may introduce time-reversal symmetric exchange anisotropies into the originally rotation invariant Haldane model in order to resolve all local degeneracies and obtain a localizable phase. The static phases are hence characterized by a single $\Z_2$ invariant corresponding to $\T_\text{edge}^2=\pm 1$. 

Given our experience with fermionic systems, it is natural to expect that the classification of driven spin-chains includes an extra topological phase that exhibits edge-spins that transform projectively under the combination of time-reversal and Floquet-time-evolution: $\T F\T^{-1}=-F^{-1}$, (in addition to the static SPT phases). This phase can be viewed as a dynamical analog of the Haldane phase, in which the spin-1/2 edge states flip under each cycle of the Floquet drive, and hence require two periods to return to their original state. This edge-state flipping can be viewed as spin-echo procedure that dynamically decouples the edge spins from bulk excitations with perfect (topologically protected) fidelity.

This suggests that the Floquet phases permit an additional dynamical $\Z_2$ topological invariant labelling whether time-reversal commutes or anti-commutes with the Floquet operator when acting on edge spins, corresponding to a total classification of $C_F[\Z_2^T] = \Z_2\times\Z_2$.

\subsubsection{Construction I: From Fermions to Spins}
In fact, we have already encountered phases with these precise realizations of all of these $\Z_2\times \Z_2$ Floquet SPT invariants in the fermionic systems with spinless time-reversal symmetry (class BDI) described above: namely the interacting version of the phase with $(N_0,N_\pi)=(2,-2)$ realizes the non-trivial Floquet invariant, and the $(4,0)$ phase realizes the nontrivial static invariant. However, the fermionic nature of this problem is unimportant, since the fermion parity ``symmetry" plays no role in the projective action of symmetry on the edge states. Hence, by adding strong interactions among the fermions, we may reduce this fermionic system to a strongly localized Mott insulator with trivially localized fermionic excitations both in the bulk and at the edges, without changing the underlying SPT order. An explicit example of an interaction term that accomplishes this task, for the $(2,-2)$ phase, in the notation of Sec.~\ref{sec:Noninteracting} is: $\sum_j U_j a_{1,j}a_{2,j}\alpha_{1,j}\alpha_{2,j}+b_{1,j}b_{2,j}\beta_{1,j}\beta_{2,j}$.

\subsubsection{Construction II: Spin-Chain}
We can also explicitly construct the phase with non-trivial static and Floquet SPT invariants directly in a purely bosonic model with spin-1 degrees of freedom. As with the AKLT construction for the ground-state SPT phase\cite{AKLT}, it is useful to construct the Floquet drive in two stage procedure where we first view each spin-1 degree of freedom, $\v{S}_i$ as being formed from two notional spin-1/2 degrees of freedom, $\sigma_i,\tau_i$, which each transform projectively under time-reversal: $\T = \prod_j\sigma_j^y\tau_j^y K$. The construction of the FLoquet SPT phase is simple in spin-1/2 description  (e.g. for the Haldane phase, consists of just nearest neighbor projectors onto singlets), and then yields a local Hamiltonian for the original non-projective degrees of freedom upon applying a local projection onto the original spin-1 degrees of freedom.

In the notional spin-1/2 language, the desired Floquet SPT phase can again be achieved by a two-step stroboscopic Floquet evolution $F=F_2F_1$, where
\begin{align}
F_1 &= e^{-iH_\text{AKLT}}
\nonumber\\
F_2 & = e^{i\pi/2 \sum_j \sigma_{A,j}^x\sigma_{B,j}^x} = i^L \prod_{j=1}^L \sigma_{A,j}^x\sigma_{B,j}^x
\end{align}
where $H_\text{AKLT} = \sum_{j=1}^{L-1} \boldsymbol{\sigma}_{B,j}\cdot \boldsymbol{\sigma}_{A,j+1}$ is the AKLT Hamiltonian whose eigenstates exhibit the SPT order of the Haldane phase.

In precisely the same manner as for the fermionic models described above, we may rewrite (dropping an irrelevant overall phase) $F_2 = \sigma^x_{A,1}\sigma^x_{B,L}\times e^{i\pi/2 \sum_{j=1}^{L-1} \sigma^x_{B,j}\sigma^x_{A,j+1}}$,  from which one may readily verify that the full Floquet evolution operator reads:
\begin{align}
F = F_2F_1 &= \sigma^x_{A,1}\sigma^x_{B,L} e^{-i\tilde{H}_\text{AKLT}}
\label{eq:FHaldane}
\end{align}
where $\tilde{H}_\text{AKLT} = \sum_{j=1}^{L-1} \lambda\(\sigma^x_{B,j}\sigma^x_{A,j+1} +\sigma^y_{B,j}\sigma^y_{A,j+1} \)+\(\lambda-\frac{\pi}{2}\)\sigma^x_{B,j}\sigma^x_{A,j+1}$ is an anisotropic analog of the AKLT Hamiltonian.

As for the Haldane phase, the edge spins: $\sigma_{A,1}$ and $\sigma_{B,L}$ are left out of $\tilde{H}_\text{AKLT}$, and hence their only dynamics is set by the preceding $\sigma^x_1\tau^x_L$ factor. In the $\sigma^z$ basis, the edge spins flip from up to down over the course of each Floquet period, producing the desired projective edge-realization of $F_\text{edge} = \sigma^x$, $\T_\text{edge} = i\sigma^y K$, such that $\(\T F\T^{-1}F\)_\text{edge}=-1$.  

From this model defined in terms of notional spin-1/2 degrees of freedom, we may obtain a corresponding Floquet evolution in terms of the original spin-1 degrees of freedom, $\v{S}_j$, by projection onto the triplet sector of each site. The terms of the AKLT Hamiltonian that appear in $F_1$ project to: $H_\text{AKLT}\rightarrow \sum_j \v{S}_j\cdot\v{S}_{j+1}+\frac{1}{3}\(\v{S}_j\cdot\v{S}_{j+1}\)^2$. The projection of the terms, $\sigma_i^x\tau_i^x$ in $F_2$, onto the spin-1 Hilbert space: $\frac{1}{2}\(\sigma_{A,i}^x+\sigma_{B,i}^x\)\rightarrow S_i^x$ can be obtained by rewriting: $\sigma_{A,i}^x\sigma_{B,i}^x = \frac{1}{2}\[\(\sigma_{A,i}^x+\sigma_{B,i}^x\)^2-2\]$, such that the Floquet drive in terms of the spin-1 variables reads $F=F_2F_1$ where:
\begin{align} 
F_1 &= e^{-i\sum_j \v{S}_j\cdot\v{S}_{j+1}+\frac{1}{3}\(\v{S}_j\cdot\v{S}_{j+1}\)^2}
\nonumber\\
F_2 & = e^{-i\pi \sum_j \(S^x_j\)^2}
\end{align}
To obtain a stable many-body localized phase, we may make the exchange couplings random, and further introduce spin-exchange anisotropies: $\lambda_j \v{S}_j\cdot\v{S}_{j+1}\rightarrow \sum_{\alpha = x,y,z} \lambda_{j,\alpha}S^\alpha_jS^\alpha_{j+1}$ to remove any unwanted local degeneracies due to continuous spin-rotation symmetry.

We note further, that the last step of projection onto a spin-1 degree of freedom is not strictly necessary to demonstrate a proof of principle construction of the SPT phase. Rather, we may instead view the model defined in terms of $\sigma_{A/B,i}$ as a complete lattice Floquet Hamiltonian for 4-state quantum degrees of freedom. In the following sections, we will hence drop the superfluous projection step.

\subsection{$\Z_n$ symmetry}
In the previous section, we considered Floquet analogs of time-reversal protected bosonic (spin) SPTs. We may also consider bosonic Floquet SPTs protected by unitary on-site symmetries. In this case, to be achieve a localized phase that is stable against heating, we may only consider Abelian symmetry groups. If we further restrict to finite-Abelian groups, than the most general symmetry group may be represented by factors of $\Z_{n_1}\times Z_{n_2}\times\dots \Z_{n_p}$ for integers $n_{1,\dots,p}\in \Z$. A prototype for this general case is to just consider a single unitary $G=\Z_n$ symmetry.

We will explicitly construct models that realize all of the projective realization of $\Z_n\times \Z$, further supporting the hypothesized classification of general interacting Floquet SPTs.
In this spirit, we can first consider an AKLT-like model in which each site is a $N^2$-state quantum with states $|m_j\>$, with $m=1\dots N^2$, that can be viewed as a tensor product of two $N$-state $\Z_N$ ``rotors", $|m_j\> = |m_{A,j}m_{B,j}\>$ labeled by sub-lattice labels $A,B$, and defined to be eigenstates of the generator, $g$, of $\Z_N$: $g|m_{A,j}m_{B,j}\> = \varphi^{m_{A,j}+m_{B,j}}|m_{A,j}m_{B,j}\>$, where $\varphi = e^{2\pi i /N}$, and $m_{A/B}\in \{0\dots N-1\}$. We can also write the (unitary) cyclical raising and lower operators: $\sigma^{\pm} = \sum_{m=0}^{N-1}|m\pm 1\mod N\>\<m|$. The $\Z_n$-symmetry generator is: $g = \prod_{j=1}^L g_{A,j}g_{B,j}$.

Then, we may realize a non-trivial Floquet SPT phase by considering the stroboscopic Floquet operator $F = F_2F_1$, with $F_1 = e^{-i\(\sum_{j=1}^{L-1}g_{B,j}g_{A,j+1}+h.c.\)}$, and $F_2 = \prod_{j=1}^L \sigma^+_{A,j}\sigma^-_{B,j} = \sigma^+_{A,0}\(\prod_{j=1}^{L-1}\sigma^-_{B,j}\sigma^+_{A,j+1}\)\sigma^-_{B,L} \equiv \sigma^+_{A,0}\sigma^-_{B,L}W$. 

Note that $W$ commutes with $F_1$, gives non-trivial phases to all bulk degrees of freedom, and does not involve sub-sites $A_1$ or $B_L$. On the ends, $g$ acts like $g_{A,1}$ and $g_{B,L}$ respectively, and $F$ acts like $\sigma^+_{A,1}$ and $\sigma^-_{B,L}$ respectively. Hence we see that on, say the left end, $F^\dagger_Lg_LF_Lg^\dagger_L = e^{2\pi i/N}$, time-translation and the $\Z_N$ symmetry are represented projectively. Moreover, we see that the $\pi$ modes of the Fermionic models are generalized to quasi-energy $\frac{2\pi}{N}$ modes for generic $n$, where the $\Z_N$ symmetry protects the enhanced. As for the AKLT chain, this projective action is preserved under the local projection of the each two-spin ``site" onto the degrees of freedom of a single non-projective $\Z_N$ spin.

Moreover, we can consider a sequence of related phases with $F_2 = \prod_{j=1}^L \(\sigma^+_{A,j}\sigma^-_{B,j}\)^n$, for $n=0,1\dots N-1$, which result in bosonic edge modes with quasi-energy fixed at $e^{2\pi n/N}$, protected by a projective interplay of Floquet-evolution and $Z_N$ symmetry at the edge: $F^\dagger_Lg_LF_Lg^\dagger_L = e^{2\pi in/N}$. These phases exhaust all projective representations of the group $\Z_N\times\Z$, in agreement with the conjectured classification of 1D bosonic Floquet SPTs: $C_F[\Z_N]=\mathcal{H}^2(\Z_N\times\Z) = \Z_N$.

\renewcommand{\arraystretch}{1.5}
\begin{table}
\begin{tabular}{C{1.1in}C{1.1in}C{1.1in}}
\toprule
{\bf Symmetry group (G)} &  {\bf Static Classification ($C[G]$)} & {\bf Floquet Classification $(C_F[G])$)} \\ 
\midrule
{\centering None} &  {\centering None} & {\centering None} \\ 
$\Z_2^T$ & $\Z_2$ & $\Z_2\times \Z_2$ \\
$\Z_n$ & {\centering None} & $\Z_n$ \\
$\Z_{n_1}\times \Z_{n_2}\times\dots \Z_{n_p}$ & $\displaystyle{\prod_{i\neq j=1}^p}Z_{\text{gcd}(n_i,n_j)}$ &  $C[G]\times \displaystyle{\prod_{i=1}^p}Z_{\text{gcd}(n_i)}$ \\
\bottomrule
\end{tabular}
\caption{{\bf Classification of 1D bosonic Floquet SPTs -- } Group structure of nontrivial topological classes for 1D bosonic systems with discrete, Abelian on-site symmetries. Non-Abelian symmetry groups and symmetry groups with anti-unitary symmetries with irreducible representations of dimension larger than one do not permit symmetry preserving many-body localization and are unstable to heating. The last entry represents the most general finite Abelian symmetry group, a derivation of the Floquet classification for this general case is present in Appendix~\ref{app:Kunneth}\label{tab:BosonicPhases}}
\end{table}

\subsection{Generalizations}
We can repeat the above construction for $\Z_N$ symmetry bosonic Floquet SPTs in more general terms for an arbitrary. In analogy to the AKLT construction, for a given projective representation, $\mathcal{PR}$, of $\Z \times G$, let us construct a model whose physical sites are composite sites of an $A$ sub-site degree of freedom (DOF) that transforms under $\mathcal{PR}$ and a $B$ sub-site DOF that transforms under the conjugate representation $\overline{\mathcal{PR}}$ (where the projective phases are complex conjugates of those in $\mathcal{PR}$). Specifically, denote the generators of elements in $G$ by $U_{A/B}(g)$, and of the the time-translation operator for the end states as $F_{A/B}$. Then, consider the case where $F_{A}=F_{B}^\dagger\equiv \mathcal{F}$, and $A$ sites transform under representation $R$ of the symmetry, and $B$ sites under it's conjugate representation $\overline{R}$ (i.e. $U_A(g) = U^\dagger_B(g) \equiv U(g)$) such that the total site (with A and B components) transforms under a non-projective representation of $\Z\times G$, but the end-states transform under conjugate projective rep's.

Then define the Floquet operator by the stroboscopic evolution $F= F_2F_1$ with: $F_2 = \prod_{j=1}^L \mathcal{F}^{\vphantom\dagger}_{A,j}\mathcal{F}^\dagger_{B,j}$, and:
\begin{align}
F_1 =\exp\[-i \sum_{\text{irreps},I}\lambda_I \sum_{j=1}^{L-1} P^I_{B,j; A,j+1}\]
\end{align}
where $P^I_{i;j}$ is the projection operator of sites $i$ and $j$ onto the $I^\text{th}$ irreducible representation (irrep), $R^I$, of $G$. If the irreps of $G$ are all singlets (i.e. have dimension one), than $F_1$ gives a different random quasi-energy to all bulk degrees of freedom resulting in many-body localization. This construction fails for non-Abelian groups with irreps of dimension higher than one, for which there are extensive local degeneracies in the quasi-energy spectrum of $F_1$. Below, we will show that this obstacle is fundamental, and that MBL is possible only for Abelian groups with irreps of dimension $1$. Hence, this classification works for all relevant symmetry groups. 

Since $U_{B,j}(g)U^\dagger_{A,j+1}$ can be block diagonalized in $R^I$, and commutes with $\mathcal{F}_{B,j}\mathcal{F}^\dagger_{A,j+1}$ (since $A$ and $B$ transform under conjugate projective representations of $\Z\times G$), such projectors will commute with $F_2$ terms in the Floquet operator.

As with the case for $\Z_N$ symmetry above, this construction results in a projective implementation of $G\times \Z$ at the edge. Namely, at the left edge, symmetry acts like $g_{A,1}$ and Floquet time evolution acts like $U_{A,j}$, which by construction satisfy a projective realization of $G\times \Z$. 

While this model is constructed at a highly fine-tuned point with zero correlation length, the results are robust to small perturbations that do not result in a phase transition. So long as the perturbation is sufficiently weak that Floquet eigenstates retain their area law entanglement structure (in the non-equilibrium Floquet setting, a phase transition is defined as a breakdown of the area-law entanglement structure of Floquet-eigenstates), then there is a well defined sense of the local action of symmetry on the edge states of the system, and hence form a projective local action of symmetry.\cite{Chen11,Fidkowski11,Pollmann10}
 Moreover, since projective representations are discrete, different projective representations cannot be continuously deformed into each other, and small perturbations cannot continuously alter the realized projective representation. 

\section{Formal classification}
In the previous sections, we have built a family of zero-correlation length (``fixed point") models that realize various fermionic and bosonic Floquet SPT phases, and support the hypothesis that the classification of these phases is given by projective representations of the symmetry group enhanced by an extra factor of $\Z$ to account for time-translation ``symmetry". In this section, we formalize these ideas, making extensive use of the ideas behind the related classification of equilibrium SPT ground-states.\cite{Chen11,Fidkowski11,Pollmann10}

Our strategy will be to construct a precise definition of the local action of symmetry, to sharpen the notion of projective interplay of on-site and time translation symmetries. To this end, consider a system with localized Floquet eigenstates protected by symmetry group $G$, we may construct an operator that commutes with the Floquet evolution, and has the same action as the locally applying a symmetry element, $g\in G$ on a large, but finite interval $I = [x_l,x_r]$ whose size greatly exceeds the localization length: $|x_r-x_l|\gg \xi$:
\begin{align}
g_{I} = U_{l,g}U_{r,g}\(\prod_{j\in I} g_j\) 
\end{align}
The middle term represents the symmetry operator restricted to sites within the interval. This term has exponentially small effect on the quasi-local quantum numbers $n_\alpha$ residing deep in the bulk of the interval, $I$, (as these commute with the unrestricted action of $g = \prod_i g_i$), and similarly exponentially small effect on quasi-local quantum numbers far away from $I$. On the other hand, this term strongly disturbs those quantum numbers near the boundaries of the interval , and hence, does not by itself commute with the Floquet evolution. However, we may repair the disturbance by acting with a pair of quasi-local unitary operators $U_{l,g}$ and $U_{r,g}$ that are exponentially well localized to the left- and right- ends of the interval respectively, which restore the state of the conserved DOF that were altered by $\prod_{j\in I} g_j$.

Paralleling Ref.~\cite{MBLSPT}, we can first construct explicit formal expressions for $U_{l/r}$ for the special case of strictly localized ``zero-correlation length" Floquet Hamiltonians, whose conserved quantities, $\{n_\alpha\}$ have bounded support on a finite number of sites. All of the models we have so far constructed take this form. Subsequently, we will adapt these ideas to the more generic case of only exponentially well localized Floquet operators.

\begin{widetext}
For zero-correlation length Floquet Hamiltonians, $\prod_{j\in I}g_j$ preserves all $n_\alpha$ whose support is fully contained inside $I$, or resides completely outside of $I$, and disturbs only a finite number, $N_{l}$ ($N_r$) of $n_\alpha$ on the left (right) boundary respectively.  We can divide the integrals of motion into four groups: those strictly in the interval $I$, those strictly in the complement of the interval, $I^c$, those intersecting the left boundary $\d I_l$, and those intersecting the right boundary, $\d I_r$, and compute the matrix elements:
\begin{align}
\(U_{l,g}^\dagger\)^{n'_{\beta_1}\dots n'_{\beta_{N_l}}}_{n_{\beta_1}\dots n_{\beta_{N_l}}}
\(U_{\vphantom{l}r,g}^\dagger\)^{n'_{\delta_1}\dots n'_{\delta_{N_r}}}_{n_{\delta_1}\dots n_{\delta_{N_l}}}
\equiv 
\<\{n_{\alpha_i\in I^c}\},\{n'_{\beta_i\in \d I_l}\},\{n_{\gamma_i\in I}\},\{n'_{\delta_i\in\d I_r}\}|
\prod_{j\in I}g_j 
|\{n_{\alpha_i\in I^c}\},\{n_{\beta_i\in \d I_l}\},\{n_{\gamma_i\in I}\},\{n_{\delta_i\in\d I_r}\}\>
\end{align}
which defines $U_{l/r,g}$ up to an overall phase. 
\end{widetext}

For the more generic case of exponentially well localized Floquet Hamiltonians, whose conserved quantities are quasi-local, the above construction is only approximate as all integrals of motion have some non-zero (albeit exponentially small) overlap with the boundaries of $I$. However, we may approximately break the $n_\alpha$ into the same groups by using an arbitrary cutoff to decide which $n_\alpha$ belong to the boundary regions $\d I_{l/r}$. This approximation is exponentially accurate in the number of integrals of motion $N_{l/r}$ taken to be in the boundary region, allowing for a well defined limiting procedure where we take the size of $I$ to infinity first, and then take $N_{l/r}$ to infinity. In this order of limits, the above construction becomes exact even for only exponentially well localized systems.  In practice, the approximation will become accurate once the sub-interval and boundary sizes are both taken to be much larger than the localization length $\xi$.

Having defined a precise notion of the local action of symmetry, we would also like to sharply define the local action of the Floquet drive near the ends of the interval $I$. To this end, we first note that a generic localized Floquet Hamiltonian of the form Eq.~\ref{eq:FMBL} may be deformed by a finite-depth local unitary transformation (to exponential-in-depth accuracy) to a 
simpler form for which the Floquet Hamiltonian decomposes into a sum of independent terms for each $n_\alpha$:
\begin{align}
\tilde{F} = \sum_{n_\alpha}e^{-i\sum_\alpha\lambda_\alpha(\Pi_{n_\alpha})}
\end{align}
Since such a finite-depth unitary circuit preserves the area-law structure of entanglement in the Floquet eigenstates, and hence cannot change the underlying phase (which would require a phase transition accompanied by a singularity in the entanglement entropy). Hence, we may, without loss of generality consider the Floquet Hamiltonian to decompose in this way. For such decomposable Floquet evolutions, we can divide the Floquet evolution operator into four independent pieces: $F \equiv F_{I^c}F_{\d I_l}F_{I}F_{\d I_r}$, and focus on the action at the left- and right- boundaries of $I$: $F_{\d I_{l/r}}$.

By construction, $g_{I}$ commutes with the Floquet evolution $F$, and forms a unitary representation of the symmetry group $G$ (i.e. $g_{I}g'_I = (gg')_I$). However, the quasi-local operators $U_{l,g}$ and $U_{r,g}$ need not separately form a representation, but rather need only satisfy the group composition rules up to an overall phase that cancels between the $l$ and $r$ end-points: $U_{l/r,g}U_{l/r,g'} = e^{\pm i\phi(g,g')}U_{l/r,gg'}$. Thus, the edge-operators $U_{l/r,g}$ need only form a projective representation of the symmetry group. As there are only a discrete set of such projective representations, the particular projective representation realized cannot be continuously altered by arbitrary perturbations, barring a phase transition that spoils the locality of the above constructions. Consequently, in the absence of periodic driving, such projective representations fully characterize the set of ground-state SPT phases.\cite{Chen11,Fidkowski11,Pollmann10}

In the Floquet system, we know that the entire object $g_I$ commutes with the Floquet evolution operator, $F$. However, separately, the local operators $\tilde{U}_{l/r,g} =U_{l/r,g}\prod_{j\in I\cap \d I_{l/r}}g_I$ need only commute with action of the Floquet evolution near the interval boundaries, $F_{\d I_{l/r}}$ up to a phase: $F_{\d I_{l/r}}\tilde{U}_{l/r,g}=e^{\pm i\phi(F,g)}\tilde{U}_{l/r,g}F_{\d I_{l/r}}$. Having opposite projective phases, $\pm\phi(F,g)$, for the left and right edges respectively ensures that the total operators will commute.

In this way, the local action of symmetry- and Floquet evolution near the edge of the interval $I$ is implemented projectively. To understand the group structure involved, we note that the Floquet evolution implements a unitary representation of the group of integers, $\Z$, where positive (negative) integers $N>0$ are respectively represented by forward (backward) time-evolution by $N$ periods: $F^N$ ($\(F^\dagger\)^N$). Thus, we see that together with the symmetry group action restricted to one end of the interval, say $U_{l,g}$, the Floquet evolution forms a projective representation of $G\times \Z$, confirming our hypothesized classification. Moreover, since this projective representation cannot be continuously altered by perturbations that preserve the locality of the Floquet eigenstates, the projective representations correspond to distinct dynamical phases. We also note that, when the full Floquet spectrum is localized, various Floquet eigenstates of a given system differ only by bulk excitations that do not change the projective action of symmetry at the edges, implying that all eigenstates must belong to the same Floquet SPT phase. Since the time evolution of an arbitrary initial state is governed by the Floquet eigenstates, then the Floquet SPT order is also imprinted on the dynamics starting from a non-eigenstate.

\section{Signatures in entanglement spectrum}
In this section, we describe signatures of intrinsic Floquet SPT order (i.e. Floquet SPT order which cannot occur in un-driven systems, or equivalently does not survive to the infinite frequency limit) in the entanglement spectrum of Floquet eigenstates. These arguments provide an alternative phrasing of the general classification presented in the previous section.

The entanglement spectrum of the Floquet eigenstate, $|\Psi\>$ can be obtained by performing a Schmidt decomposition: $|\Psi\> = \sum_n \frac{e^{-\eps_n/2}}{\sqrt{Z}}|\Psi_{n,L}\>|\Psi_{n,R}\>$ where $|\Psi_{L/R}\>$ are states living on the left and right of the entanglement cut respectively, such that the reduced density matrix for the left half of the system: $\rho_L = \sum_n \frac{e^{-\eps_n}}{Z}|\Psi_{n,L}\>\<\Psi_{n,L}|$, takes the form of a thermal density matrix with entanglement Hamiltonian $H = \sum_n \eps_n|\Psi_{n,L}\>\<\Psi_{n,L}|$. Here $Z =\sum_ne^{-\eps_n}$, normalizes the trace of the reduced density matrix. Note that, unlike the quasi-energy spectrum, the entanglement spectrum is non-compact and is not periodic modulo $2\pi$. Consequently, $\eps = 0$ is a special entanglement energy dividing positive and negative states, unlike quasi-energies whose absolute value has no meaning.

For equilibrium 1D SPTs, the entanglement spectrum exhibits degenerate zero modes that permit one to diagnose the SPT order. For systems with on-site symmetries, the entanglement spectrum exhibits zero-modes ($\eps_n=0$) with multiplicity equal to the edge state degeneracies of a system with open boundary conditions.\cite{Pollmann10,Fidkowski10} For 1D crystalline SPTs (e.g. protected by inversion), the entanglement spectrum contains zero-modes indicative of the SPT order, even in cases where a physical edge would break the protecting symmetry and fail to exhibit edge states.\cite{Pollmann10}. Acting with the symmetry operations on either side of the entanglement cut reveals the projective action of symmetry on the edge states. This bulk-edge correspondence provides a numerically testable probe of the equilibrium SPT order in a system without boundaries.

However, for a state with intrinsic Floquet SPT order but no equilibrium SPT order, i.e. projective edge action of $F$ and $G$, but non-projective edge action of symmetry alone, there is no universal signature in the static entanglement spectrum of a Floquet eigenstate. This is manifestly seen by considering the special case of a zero correlation Floquet drive that realizes the FSPT order with $\Z_2$ symmetry. Rather than considering the AKLT like model described above, which allows us to construct states with either equilibrium or driven SPT order on the same footing, we may consider the simpler model introduced in Ref.~\cite{Khemani15}, that realizes the nontrivial FSPT phase. This model consists of a spin-1/2 chain with Ising symmetry generated by $g = \prod_i \sigma^x_i$, and stroboscopic Floquet drive:
\begin{align}
F_{\Z_2} &= e^{-i\pi/2\sum_{i=1}^L \sigma^z_{i}\sigma^z_{i+1}}e^{-i\sum_{i=1^L} h_i\sigma^x_i} 
\nonumber\\
&= \tilde{\sigma}^z_{1}\tilde{\sigma}^z_L e^{-i\sum_{i=2}^{L-1} h_i\sigma^x_i}
\label{eq:FZ2}
\end{align}
where $\tilde{\sigma}^z_{1/L}=\(e^{-ih_{1/L}/2\sigma^x_{1/L}}\sigma^z_{1/L}e^{ih_{1/L}/2\sigma^x_{1/L}}\)$ are rotated Pauli matrices with a quantization access tilted along the $\cos(h_{1/L})\hat{z}+\sin(h_{1/L})\hat{y}$ direction in the xy-plane. We can readily verify that $F_{\Z_2}$ realizes the projective edge action of $\{F,g\}_\text{edge}=0$, since $F_\text{edge} = \tilde\sigma^z$ and $g_\text{edge} = \sigma^x = \tilde\sigma^x$, which anti-commute. Moreover, since $F_{\Z_2}$ is a product of on-site unitary operators, its eigenstates are just product states, and hence have trivial entanglement spectrum.

Thus the static entanglement spectrum does not contain information about the topological edge states, unlike the equilibrium case. However, the SPT order does manifest itself if we examine the full time-dependent micro-motion of the entanglement spectrum Floquet eigenstates for times $0\leq t\leq T$. To see this, let us continue working with the special zero-correlation length Hamiltonian of Eq.~\ref{eq:FZ2}. Let us consider the entanglement spectrum of a particular Floquet eigenstates, $|\Psi\>=\otimes_i|s_i\>$, where $|s_i\>$ are $\sigma^x_i$ eigenstates with eigenvalue $s_i = \pm 1$. The first phase of the Floquet evolution, $U(t,0)=e^{-i t/T_1\sum_i h_i  \sigma^x_i}$, just generates an overall phase for $|\Psi\>$, and does not effect the entanglement spectrum. The second stage: $U_2(t+T_1,T_1) = e^{-it/T_2\sum_i \sigma^z_i\sigma^z_{i+1}} = U_{2,L}U_{2,R}U_{2,\text{cut}}$ can be decomposed into pieces that act only on the left and right half, and one term that acts across the cut: $U_2 = U_{2,L}U_{2,R}U_{2,\text{cut}}$, with $U_{2,L} = e^{-i\pi t/2T_2\sum_{i\leq -1}\sigma^z_i\sigma^z_{i+1}}$, $U_{2,R} = e^{-i\pi t/2T_2\sum_{i\geq 1}\sigma^z_i\sigma^z_{i+1}}$, and $U_{2,\text{cut}} = e^{-i\pi t/2T_2\sigma^z_0\sigma^z_{1}}$. Since only $U_{2,\text{cut}}$ generates entanglement, we may equivalently consider the simplified problem of finding the entanglement spectrum of the two spin system straddling the cut, evolving according to $U_{2,\text{cut}}$. Explicitly, we have:
$e^{-i\pi t/2T_2\sigma^z_0\sigma^z_{1}}|s_0s_1\> = \cos\(\frac{\pi t}{2T_2}\)|s_0s_1\> +\sin\(\frac{\pi t}{2T_2}\)|-s_0,-s_1\>$
i.e. the reduced density matrix of the left side is: $\rho_L(t + T_1) = \cos^2\(\frac{\pi t}{2T_2}\) |s_0\>\<s_0|+\sin^2\(\frac{\pi t}{2T_2}\)|-s_0\>\<-s_0| = \frac{e^{-h(t)\sigma^x_0}}{Z}$, where: $h(t) = \mp \tanh^{-1}\[\cos\(\frac{\pi t}{T_2}\)\]$ for $s_0=\pm 1$.  

\begin{figure}[tb]
\begin{center}
\includegraphics[width=0.7\linewidth]{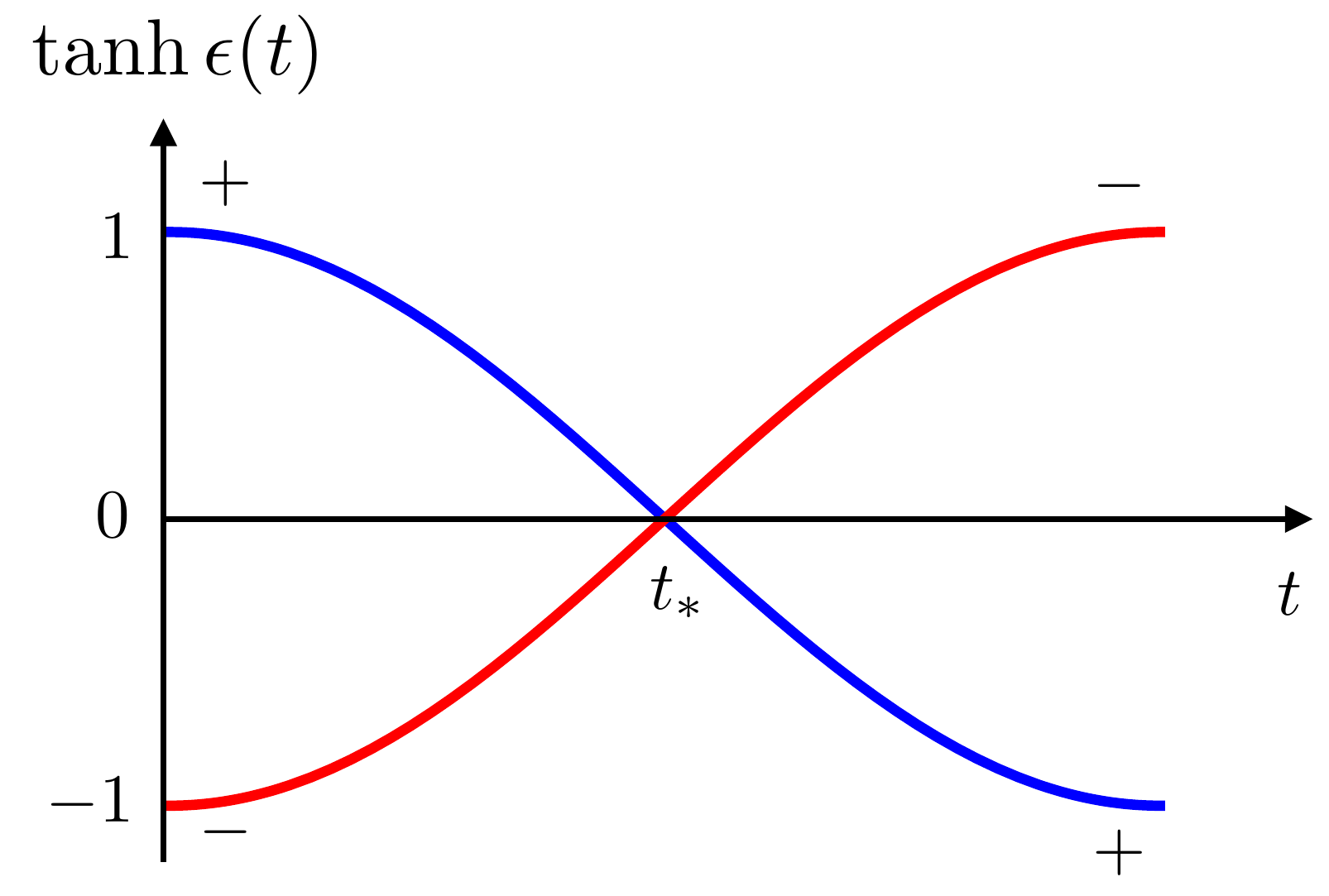}
\end{center}
\vspace{-0.2in}
\caption{
{\bf Symmetry-charge pumping in micro-motion of entanglement spectrum} - The quasi-energy spectrum (re-parameterized by $\tanh$ to fall between $\pm 1$ rather than $\pm \infty$), of the $\Z_2$-symmetry protected bosonic FSPT exhibits a quantized pumping of $\Z_2$ symmetry charge, in each Floquet cycle.
}
\label{fig:pumping}
\end{figure}

We see that the entanglement spectrum contains two eigenvalues: $\eps= \pm h(t)$ (see Fig.~\ref{fig:pumping}), whose corresponding Schmidt states have opposite $\Z_2$ eigenvalue, and which are initially at $\pm \infty$ at the beginning of the Floquet period ($t=0$). During the second stage of the Floquet evolution, $h(t)$ changes decreases $+\infty$ towards $-\infty$, crossing zero at time $t_* = T_1+T_2/2$, at which point the entanglement spectrum becomes degenerate and the two $|\pm s_0\>$ branches cross each other. Continuing the evolution, the entanglement spectrum returns towards $\eps = \pm \infty$, but with the $|\pm s_0\>$ branches exchanged. Fixing $s_0=+1$ for concreteness, we see that the $\Z_2$ symmetry charge of the negative entanglement energy bands ($\eps<0$) changes by one unit, as the $|- \>$ branch of the spectrum exchanges with the $|+ \>$ branch. This pumping of symmetry charge provides a bulk probe of the projective action of Floquet evolution and symmetry at the edge ($(FgF^{-1}g)_\text{edge}=-1$), as we will explain in more detail below. We note that taking into account $U_{2,L}$ does not affect this pumping of symmetry charge, because $U_{2,L}$ commutes with $U_{2,\text{cut}}$ and the symmetry operation $g$ restricted to the left half chain. Note also that, to diagnose the Floquet SPT order we need the full micro-motion of the entanglement spectrum, rather than just the spectrum at any single time-cut. 

While we have worked out this structure for a particular example, the pumping of symmetry charge in the entanglement spectrum turns out to be a robust way to characterize the Floquet SPT order, and is equivalent to the projective action of symmetry and time-translation obtained at the edges of a finite system with open boundaries. To see this, first note that since the $t=0$ and $t=T$ end-points of the evolution have gapped, zero-dimensional entanglement spectra, we may classify the SPT properties of the entanglement ``ground-state" (i.e. of all ``occupied" Schmidt-states with $\eps <0$). For a bosonic zero-dimensional system with symmetry $G$, the SPT invariant is given by the group of one-dimensional representations of the group, $\mathcal{H}_1\(G,U(1)\)$, i.e. the ``symmetry charge" of the entanglement ground-state. In the above example, we have seen that, for a nontrivial intrinsically Floquet SPT there is a robust change in the symmetry charge of the entanglement ground state over the course of one cycle. This difference between initial and final entanglement symmetry charges cannot be changed without closing the entanglement gap at $t=0,T$, i.e. by driving a phase transition into a different phase, and is hence a robust characteristic of the FSPT phase. Moreover, the group of such one-dimensional group representations coincides exactly with the extra factor of $\mathcal{H}_1\(G,U(1)\)$ appearing in the Kunneth formula for classifications of $G\times \Z$, and hence agrees precisely with the interpretation of 1D FSPT phases as having projective action of edge symmetries, but provides an alternative perspective that is testable in systems without boundaries. A closely related picture holds for the fermionic FSPT states, though here fermion parity plays a key role in the pumping (e.g. one must keep track of the pumping of fermion parity and symmetry charges across the entanglement cut).

\section{Symmetry constraints on localizability}
Localization is crucial to avoid heating and obtain Floquet eigenstates with area-law entanglement entropy, which permits sharp distinctions between dynamical phases -- e.g. with entanglement entropy serving as a free energy, whose singularities in the limit of infinite system size represent phase transitions. In this section, we show that the requirement of localization places strong constraints on the type of symmetry groups that we may consider. We will derive a general criterion based on the representation theory of the symmetry group, namely, that symmetry-preserving localization is only possible for groups whose irreducible representations (irreps) all have dimension one -- i.e. only for Abelian symmetry groups. This constraint applies to both static MBL Hamiltonians, and periodic Floquet MBL systems alike.

In a symmetry preserving MBL system, the symmetry generators $g$ must commute with the projection operators $\Pi_{n_\alpha}$ onto the states of the local conserved quantities.  Thus the different values of $n_\alpha$ label one-dimensional representations of $G$. For a generic symmetric Hamiltonian, the different values of $n_\alpha$ will label irreps of $G$ (as reducible representations can be sub-divided into irreps by the application of a infinitesimal local perturbations). Moreover, since the microscopic degrees of freedom must form a faithful representation of the symmetry group (otherwise the true symmetry group should be regarded as a subset of $G$), tensor products of lattice-scale degrees of freedom will generate all possible irreps of $G$.

When $G$ is Abelian, all irreps are one-dimensional and the state labelled by $|n_1n_2\dots n_L\>$ is unique, and well-defined. On the other hand, when $G$ is non-Abelian, some values of $n_\alpha$ necessarily correspond to irreps with dimension $D_{n_\alpha}>1$.
In this case, the quantum numbers $n_\alpha$ can at most label local irreps of the symmetry group, each of which must be augmented with some additional quantum numbers $q_\alpha = 1\dots D_{n_\alpha}$ to specify a quantum state. In a generic state, local excitations that transform under such multidimensional irreps will be present at finite density in a generic state: such that ``$|n_1n_2\dots n_L\>$" actually corresponds to a collection of extensively degenerate $\prod_\alpha D_{n_\alpha}\sim e^{AL}$ states $|(n_1,q_1);(n_2,q_2)\dots\rangle$ for some constant $A = \sum_{\text{irreps},I}\log D_I \rho_I$, where $\rho_I$ is the density of excitations in the $I^\text{th}$ irrep.

Such an extensive degeneracy will be inherently unstable to arbitrarily small perturbations, which will lead to interactions and among the locally degenerate excitations, resulting in resonant quantum fluctuations that will resolve the extensive local degeneracy. However, regardless of the details of this degeneracy lifting by fluctuations, there is no possible localized state that respects the symmetry. Instead, we see three conceivable alternative outcomes:
\begin{enumerate}
\item Quantum fluctuations among the highly degenerate states can lead to thermalization and a break down of MBL.
\item The state may spontaneously lift the degeneracy by choosing a product states of quantum numbers $q_\alpha$, however, this necessarily corresponds to a spontaneous breaking of symmetry $G$ down to an Abelian sub-group, since the auxiliary quantum numbers $q_\alpha$ transform non-trivially under $G$.
\item If the residual interactions between pairs of $(n_\alpha,q_\alpha)$ with $D_{n_\alpha}>1$ are strongly random and local, the system may form a quantum critical state that is neither thermal, nor strictly localized. This state can be viewed as a generalized random singlet phase, such as those recently identified in loosely related systems of random anyonic chains.\cite{Vasseur15QCG}
\end{enumerate}
Options $1$ and $2$ were both recently observed in renormalization group and numerical studies\cite{Vasseur15XXZ} of 1D topological chains of fermions with random hopping, whose ground-state forms an SPT protected by $G = U(1)\times \Z_2^T$, where $\Z_2^T$ corresponds to anti-unitary time-reversal symmetry. For this system, there is one singlet (with zero $U(1)$ charge), and  an infinite number of $D=2$ irreps with integer non-zero $U(1)$ charge $\pm n$, that are interchanged by time-reversal. At weak disorder, the symmetry-ensured local degeneracies correspond to strongly overlapping degrees of freedom that lead to thermalization. At strong disorder, the excited states of this model were found to inevitably spontaneously break the $\Z_2^T$ reversal. In the strong disorder renormalization group treatment, this spontaneous symmetry breaking arises due to the accumulation of clusters with increasingly large charge $q$, strongly suppressing quantum fluctuations, and leaving essentially dominantly classical interactions that lead to symmetry breaking. The strong disorder physics of this model, is potentially special to the presence of an infinite number of irreps. The third, critical option described above likely is only a possibility for discrete non-Abelian groups, with a finite number of irreps, however we leave the precise properties of this for future study.

While we presented results for unitary symmetry groups, we note that this constructions can be readily generalized to anti-unitary time-reversal symmetry by using the ideas of Ref.~\cite{Chen15} to define the local action of complex conjugation on tensor-product states. For example, in a fermionic system, there must be local fermionic excitations, i.e. different values of $n_\alpha$ must label states with either even- or odd- local fermion parity. In a system where time-reversal squares to $(-1)$ in the odd fermion sector ($\T^2=P_f$), there will be a local Kramers degeneracy in the odd local fermion parity sectors of each subsystem $\alpha$. Consequently, these arguments also rule out MBL SPT phases protected by time-reversal symmetry with degrees of freedom with Kramers doublet fermions ($\T^2=P_f$), such as the familiar 2D and 3D electronic time-reversal symmetric topological insulator materials realized in solid-state materials. 

\section{Discussion}
We have shown that the classification of topological phases in 1D Floquet systems, can be understood by generalizing the equilibrium classification to include an extra time-translation symmetry, and have provided explicit model constructions of a large class of Floquet SPT phases. A simple generalization of our arguments to higher dimension, $d$, would suggest that the bosonic Floquet SPT classification with symmetry group $G$ is given by higher cohomology groups: $\mathcal{H}^{d+1}\(\tilde{G},U(1)\)$, where again $\tilde{G}$ consists of $G$ enhanced by time-translation symmetry (e.g. $\tilde{G} = G\times \Z$ for unitary $G$ or $G\rtimes \Z$ for anti-unitary $G$ as appropriate).\cite{CohomologyScience,CohomologyPRB} Characterizing the phenomenology of these phases, and understanding the classification of higher-dimensional interacting fermion phases presents interesting challenges for future work.

Having obtained a systematic theoretical understanding of the structure of topological phases in 1D Floquet systems, a natural next step will be to investigate potential experimental realizations of these phases in cold-atom or other quantum optics based systems. To this end, the most promising candidate seems to be spin-systems, such as the dynamical analog of the Haldane chain, since the non-trivial fermionic phases do not permit localization, either due to the non-Abelian nature of their symmetry group, or because they occur in explicitly particle number non-conserving systems, i.e. superfluids, which in cold-atoms contexts possess extended Goldstone modes that will act as a thermalizing bath. 

If realized, the Floquet Haldane phase may be diagnosed experimentally by the absence of decoherence for the edge-spins. Namely, the timescale for decoherence of an initially prepared quantum state of the Floquet topological edge states, comes only from the interaction between edge states on opposite sides of the system, and the decoherence time diverges exponentially in the length of the system. This coherent storage is also present for non-driven MBL topological phases, however, the Floquet topological phases may be distinguished by noting that the edge-state spin coherently flips over the coarse of each Floquet period. One can probe the symmetry protected nature of the Floquet SPT edge states by intentionally introducing a symmetry breaking field\cite{Bahri15} to induce decoherence of the edge state information, which can be subsequently reversed by applying the opposite symmetry breaking field.

\vspace{0.1in}\noindent\textit{Acknowledgements} -- We thank R Vasseur, M Serbyn for insightful discussions. We also thank , C von Keyserlingk, and S Sondhi for sharing their unpublished preprint on classifying 1D Floquet SPTs. This work was supported by by the Gordon and Betty Moore FoundationÕs EPiQS Initiative through Grant GBMF4307 (ACP and TM), a Simons Investigator Grant and NSF DMR 1411343 (AV).

\vspace{0.1in}\noindent\textit{Note -- } During the completion of this work we became aware of an related independent works by C von Keyserlingk and S. Sondhi \cite{Keyserlingk16}, and D. Else and C. Nayak\cite{Else16}, whose results are consistent with our own, where they overlap.

\appendix
\section{Bilinear couplings between a MZM and a MPM
\label{app:noninteracting}}
In this section, we show that a bilinear coupling between a MZM and a MPM does not change quasi-energies of the Majorana end states for the specific model of superconducting chain of spinless fermions subject to the stroboscopic periodic drive defined in Eqs. (\ref{eq:HMPM}). 
We consider two superconducting chains denoted by Majorana fermions
$a_i, b_i$ and $a'_i, b'_i$ that support a MPM and a MZM, respectively.
Specifically, we focus on the Hamiltonian given by:
\begin{align}
&H(t) = \nonumber \\
&	\begin{cases}
	H_1  = \frac{i\lambda_1}{4}\sum_{j=1}^L (a_jb_j +a'_j b'_j), & 0\leq t < T_1, \\[6pt]
	H_2  = \frac{i\lambda_2}{4}\sum_{j=1}^{L-1} (b_j a_{j+1} + b'_j a'_{j+1}), & T_1\leq t < T_1+T_2,
	\end{cases}
	\label{eq:HMPM}
\end{align}
with $\lambda_1 T_1/4=\pi/2$.
In this case, the Floquet operator reads
\begin{align}
F&=a_1 F' b_L, \\
F'&= e^{-i \tilde T_2 H_2} \prod_{j=1}^{L-1} a'_j b'_j .
\end{align}
Here we note that $F'$ commutes with $a_1$ and $a'_1$.
Since $F a_1 F^\dagger=-a_1$ and $F a'_1 F^\dagger=a'_1$ hold,
$a_1$ and $a'_1$ are a MPM and a MZM, respectively.

Now we add a coupling between the MPM and the MZM by adding the bilinear term $i(\delta/T_2) a_1 a'_1$ term to $H_2$.
This modifies the Floquet operator as
\begin{align}
F&=a_1 e^{-\delta a_1 a'_1} F' b_L,
\end{align}
and $a_1$ and $a'_1$ no longer describes eigenstates of $F$;
the operator $a_1$ ($a'_1$) does not satisfy 
$F a_1 F^\dagger = \epsilon a_1$ ($F a_1 F^\dagger = \epsilon' a_1$)
with quasienergy $\epsilon$ ($\epsilon'$).
Instead, eigenstates are given by superpositions of $a_1$ and $a'_1$ as
\begin{align}
\tilde a_1 &= a_1 e^{-\delta a_1 a'_1} = a_1 \cos \delta - a'_1 \sin \delta, \\
\tilde a'_1 &= a'_1 e^{-\delta a_1 a'_1} = a_1 \sin \delta + a'_1 \cos \delta.
\end{align}
These Majorana fermions $\tilde a_1$ and $\tilde a'_1$ satisfy
\begin{align}
F\tilde a_1 F^\dagger&=-\tilde a_1,
&
F\tilde a'_1 F^\dagger&=\tilde a'_1,
\end{align}
and corresponds to a new MPM and a MZM, respectively.
Thus the bilinear coupling for a MPM and a MZM does not change the quasi-energies. It only modifies associated Majorana operators.

\begin{widetext}
\section{Derivation of $H^2(\prod_i \Z_{n_i}, U(1))$ from the Kunneth formula \label{app:Kunneth}}
In this section, we derive the second cohomology group $H^2(\prod_i \Z_{n_i}, U(1))$ that appears in Table \ref{tab:BosonicPhases}.
This involves the universal coefficient theorem and the Kunneth formula for cohomology and homology groups \cite{Hatcher02,CohomologyPRB,CohomologyScience,Morimoto14}. First, the universal coefficient theorem relates the cohomology group to the homology group as:

\begin{align}
H^2 \left(\prod_i \Z_{n_i}, U(1) \right)=H_2 \left(\prod_i \Z_{n_i}, \Z \right).
\end{align}
For this homology group, we apply the Kunneth formula,
\begin{align}
H_2(G_1 \times G_2, \mathbb{Z}) &=
\prod_{i=0}^2 H^i(G_1,\mathbb{Z}) \otimes H^{2-i}(G_2,\mathbb{Z})
\nonumber \\
&\qquad 
\times \prod_{i=0}^1 \mathrm{Tor}_1^\mathbb{Z}[H^i(G_1,\mathbb{Z}) , H^{1-i}(G_2,\mathbb{Z})].
\label{eq: kunneth}
\end{align}
by using the following equations\cite{CohomologyPRB}:
\begin{align}
\mathbb{Z}_{n_1} \otimes \mathbb{Z}_{n_2} &= \mathbb{Z}_{\mathrm{gcd}(n_1,n_2)}, \\
G_1 \otimes (G_2 \times G_3) &= (G_1 \otimes G_2) \times (G_1 \otimes G_3), \\
H^0(\Z_n,U(1))&=H_0(\Z_n,\mathbb{Z})=\mathbb{Z}, \\
H^1 \left(\prod_i \Z_{n_i},U(1)\right)&=H_1 \left(\prod_i \Z_{n_i},\mathbb{Z}\right)=\prod_i \mathbb{Z}_{n_i}, \\
H^2(\Z_n,U(1))&=H_2(\Z_n,\mathbb{Z})=0.
\end{align}

Now the second cohomology group $H^2(\prod_i \Z_{n_i}, U(1))$ is obtained by successively applying the Kunneth formula as:
\begin{align}
H^2 &\left(\prod_{i=0}^p \Z_{n_i}, U(1) \right)
=H_2 \left(\prod_{i=0}^p \Z_{n_i}, \mathbb{Z} \right)
\nonumber \\ 
&=
\left[H_0(\Z_{n_1}, \mathbb{Z}) \otimes H_2 \left(\prod_{i=1}^p \Z_{n_i}, \mathbb{Z} \right) \right]
\times 
\left[H_1(\Z_{n_1}, \mathbb{Z}) \otimes H_1 \left(\prod_{i=1}^p \Z_{n_i}, \mathbb{Z} \right) \right]
\times 
\left[H_2(\Z_{n_1}, \mathbb{Z}) \otimes H_0 \left(\prod_{i=1}^p \Z_{n_i}, \mathbb{Z} \right) \right]
\nonumber \\
&=
H_2 \left(\prod_{i=1}^p \Z_{n_i}, \mathbb{Z}\right) \times \prod_{i=1}^p \mathbb{Z}_{\mathrm{gcd}(n_1,n_i)}
= \ldots 
=\prod_{i<j} \mathbb{Z}_{\mathrm{gcd}(n_i,n_j)}.
\end{align}
We note that the Tor functor part in Eq.~(\ref{eq: kunneth}) vanishes because of $\mathrm{Tor}_1^\mathbb{Z}[\mathbb{Z},\prod_i \mathbb{Z}_{n_i}]=\mathrm{Tor}_1^\mathbb{Z}[\mathbb{Z}_{n_i},\mathbb{Z}]=0$.

In a similar manner, $H^2(G\times \Z,U(1))$ with $G=\prod_{i=0}^p \Z_{n_i}$ is obtained by applying the above procedure for $G\times \Z$ and using 
$\mathbb{Z}_n \otimes \mathbb{Z}=\mathbb{Z}_n$ (crudely speaking, $\mathrm{gcd}(n,\infty)=n$) as
\begin{align}
H^2(G\times Z,U(1)) &= \left(\prod_{0\le i<j \le p} \mathbb{Z}_{\mathrm{gcd}(n_i,n_j)} \right) \times G.
\end{align}
This cohomology group gives the Floquet classification of 1D bosonic systems with the symmetry group $G=\prod_{i=0}^p \Z_{n_i}$ in Table \ref{tab:BosonicPhases}.
\newpage
\end{widetext}

\section{Projective representations of $\Z\rtimes \Z_2^T$}
In this appendix, we derive the projective representations of $\Z\rtimes \Z_2^T$, corresponding to the classification of time-reversal invariant bosonic FSPTs. The results do not follow from the Kunneth formula explained in the previous section, due to the semidirect product structure, but can be obtained directly.
\subsection{Bosonic systems}
Denoting the generator of $\Z$ by $F$ (Floquet evolution), and the generator of time-reversal, $\Z_2^T$, as $\T$, the group relations are: $\(\T^2\)_\text{group} = 1$ and $\(\T F\T^{-1}F\)_\text{group} = 1$. At the edge of a Floquet SPT these can be implemented projectively as $\T^2 = \omega_T$ and $\T F\T^{-1}F = \omega_{T,F}$, where $\omega_{T,F} \in U(1)$ are phases. Because $\T$ is antiunitary, $\omega_{T,F}$ cannot be altered by a simply redefinition of $F$ or $\T$ by an overall phase, and hence, if consistent, different values of $\omega_{T,F}$ correspond to distinct projective representations. 

The possible consistent values of $\omega_{T}$ can be identified as follows. Since $\T$ is antiunitary, associative requires $\T^3 = \T(\T^2) = \T(\omega_T) = \omega^*_T\T = (\T^2)\T = \omega_T\T$, i.e. that $\omega_T$ is real, allowing for two solutions: $\omega_T = \pm 1$. To fix the possible values of $\omega_{T,F}$, we first note that $(\T F\T^{-1})^{-1} = \T F^{-1}\T^{-1} = \omega^*_{T,F} F$. Using this relation we see that $\T^2 F\T^{-2}=\T \omega_{T,F} F^{-1} \T^{-1}=\omega^*_{T,F}\T F^{-1} \T = \(\omega^*_{T,F}\)^2F$, but, on the other hand, $\T^2 F\T^{-2} = |\omega_T|^2 F = F$. Together these relations require: $\omega_{T,F} = \pm 1$. 

Together, there are four projective representations of $\Z\rtimes\Z_2^T$ corresponding to $\omega_{T,F} = \pm 1$, corresponding to a $\Z_2\times\Z_2$ group structure.

\subsection{Fermionic systems}
For fermion systems, there is an additional $\Z_2$ fermion parity ``symmetry" $P_f$. This gives an additional pair of gauge invariant  group relations: $\(TP_fT^{-1}P_f\)_\text{group} = 1$, and $\(FP_f F^{-1}P_f^{-1}\)_\text{group} = 1$, which can be modified to projective relations $\T P_f\T^{-1}P_f = \omega_{\T,P}$ and $FP_fF^{-1}P_f = \omega_{F,P}$. Consistency between $\(\T P_f\T\)^{-1} = \T P_f \T^{-1} = \omega_{T,P}P_f$ and $\(\T P_f\T\)^{-1} = \(\omega_{T,P} P_f\)^{-1} = \omega_{T,P}^*P_f$, implies that $\omega_{T,P}$ must be real: $\omega_{T,P} = \pm 1$. Repeating the same line of reasoning with $\T\leftrightarrow P_f$, requires $\omega_{F,P}=\pm 1$. 

For just $\T$ and $P_f$ alone, there are $4$ distinct projective representations corresponding to $\omega_{T,P}=\pm 1$ and $\omega_{T,F} = \pm 1$. These correspond to the even entries of the $\Z_8$ classification of $\T^2=1$ fermions in 1D (the odd entries have unpaired Majorana zero modes, corresponding to fractional fermion parity and do not fit into the language of projective representations of symmetry). Adding the Floquet drive to the mix, gives an additional four possibilities: $\omega_{T,F}=\pm 1$, and $\omega_{P,F} = \pm 1$. These correspond to the additional factor of $\Z_4$ in the BDI classification (note that we have shown in the main text that two copies of the phase with $\omega_{P,F}=-1$ has $\omega_{T,F}=-1$).

\bibliography{FloqSPTbib}

\end{document}